 \documentclass[preprint2,twoside]{myaastex}

\usepackage{amssymb}                       
\usepackage{amsmath}                       
 \usepackage[ps2pdf,bookmarks=true,breaklinks=true,hypertexnames=false,colorlinks=true,pdfstartview=FitV,linkcolor=blue,citecolor=blue,urlcolor=blue]{hyperref}
\newcommand{\myemail}{gennaro@astro.ex.ac.uk}
\newcommand{\bmath}[1]{\boldsymbol{#1}}
\newcommand{\rmn}[1]{\mathrm{#1}}
\newcommand{\mathbfss}[1]{\boldsymbol{\mathsf{#1}}}
\newcommand{\ddt}[1]{\frac{\partial #1}{\partial t}} %
\newcommand{\dd}[2]{\frac{\partial #1}{\partial #2}} %
\newcommand{\gvec}[1]{\bmath{#1}}                 
\newcommand{\gdiv}{\nabla\bmath{\cdot}}           
\newcommand{\gOmega}{\mathbf\Omega}               
\newcommand{\gdotOmega}{\dot{\mathbf\Omega}}      
\newcommand{\vdp}{\bmath{x}-\bmath{x}_{\rmn{p}}}  
\newcommand{\mdp}{|\bmath{x}-\bmath{x}_{\rmn{p}}|}
\newcommand{\mxp}{|\bmath{x}_{\rmn{p}}|}          
\newcommand{\refEqt}[1]{Equation~(\ref{#1})}      
\newcommand{\refeqt}[1]{equation~(\ref{#1})}      
\newcommand{\refeqs}[2]{equations~(\ref{#1}) and (\ref{#2})}
\newcommand{\refeqp}[1]{eq.~[\ref{#1}]}           
\newcommand{\refFgt}[1]{Figure~\ref{#1}}          
\newcommand{\refFgp}[1]{Fig.~\ref{#1}}            
\newcommand{\refTab}[1]{Table~\ref{#1}}           
\newcommand{\refSect}[1]{Section~\ref{#1}}        
\newcommand{\refSects}[2]{Sections~\ref{#1} and \ref{#2}}
\newcommand{\refSecp}[1]{section~\ref{#1}}        
\newcommand{\MSun}{\mbox{$M_{\sun}$}}             
\newcommand{\MStar}{\mbox{$M_*$}}                 
\newcommand{\MJup}{\mbox{$M_{\mathrm{J}}$}}       
\newcommand{\Mp}{\mbox{$M_{\rmn{p}}$}}            
\newcommand{\Md}{\mbox{$M_{\rmn{D}}$}}            
\newcommand{\taum}{\tau_{\rmn{M}}}                
\newcommand{\Rhill}{\mbox{$R_\rmn{H}$}}           
\newcommand{\AU}{\mbox{\textrm{AU}}}              

\newcommand{\sdunits}{\mbox{$\rmn{g}\,\rmn{cm}^{-2}$}}
\received{year Month day}
\revised{year Month day}
\accepted{year Month day}
\ccc{code}
\cpright{none}{73 AC}

\slugcomment{\small{November 2004}}

\shorttitle{Migration rates of protoplanets}
\shortauthors{D'Angelo, Bate, \& Lubow}

\begin{document}


\title{\textbf{%
The dependence of protoplanet migration rates on coorbital torques\footnote{%
        To appear in the
        \textsc{Monthly Notices of the Royal Astronomical Society}.
        A version with high resolution figures is available at 
        \mbox{~~\hspace*{40ex}~~} 
        \url{http://www.astro.ex.ac.uk/people/gennaro/publications/}.
         }}}


\author{\textsc{Gennaro D'Angelo and Matthew R. Bate}}
\affil{School of Physics,
       University of Exeter,
       Stocker Road,
       Exeter EX4 4QL,
       United Kingdom}
\email{\myemail,
       mbate@astro.ex.ac.uk\\[2mm]}
\and
\author{\vspace*{-5mm}\textsc{Steve H. Lubow}}
\affil{Space Telescope Science Institute,
       3700 San Martin Drive, 
       Baltimore, 
       MD 21218, USA}
\email{lubow@stsci.edu\\[-5mm]}




\small

\begin{abstract}
We investigate the migration rates of high-mass protoplanets embedded 
in accretion discs via two and three-dimensional hydrodynamical
simulations. The simulations follow the planet's radial motion and
employ a nested-grid code that allows for high resolution close to
the planet. We concentrate on the possible role of the coorbital
torques in affecting migration rates. We analyse two cases: 
(\textit{a}) a Jupiter-mass planet in a low-mass disc and (\textit{b}) 
a Saturn-mass planet in a high-mass disc. The gap in case (\textit{a}) 
is much cleaner than in case (\textit{b}). Planet migration in case 
(\textit{b}) is much more susceptible to coorbital torques than in case 
(\textit{a}). We find that the coorbital torques in both cases do not 
depend sensitively on whether the planet is allowed to migrate through 
the disc or is held on a fixed orbit. 
We also examine the dependence of the planet's migration rate on the 
numerical resolution near the planet.
For case (\textit{a}), numerical convergence is relatively easy to obtain, 
even when including torques arising from deep within the planet's Hill 
sphere, since the gas mass contained within the Hill sphere is much less 
than the planet's mass. The migration rate in this case is numerically 
on order of the Type~II migration rate and much smaller than the Type~I rate, 
if the disc has $0.01$ solar-masses inside $26\,\AU$. 
Torques from within the Hill sphere provide a substantial opposing 
contribution to the migration rate.
In case (\textit{b}), the gas mass within the Hill sphere is larger than 
the planet's mass and convergence is more difficult to obtain. Torques 
arising from within the Hill sphere are strong, but nearly cancel. Any
inaccuracies in the calculation of the torques introduced by grid 
discretization can introduce spurious torques. If the torques within the 
Hill sphere are ignored, convergence is more easily achieved but 
the migration rate is artificially large.
At our highest resolution, the migration rate for case (\textit{b}) is 
much less than the Type~I rate, but somewhat larger than the Type~II rate.
\end{abstract}




\section{Introduction}
\label{sec:introduction}
When the first planetary systems were discovered, migration provided the 
natural explanation for the existence of the so-called ``Hot Jupiters'' 
\citep*{lin1996}. For this explanation to hold, 
migration time-scales should be no longer than disc life-times of several 
million years \citep*[e.g.,][]{Haisch2001}. However, the migration and planet 
formation processes are inter-related. Clearly, there would be complications 
and possibly difficulties in understanding planet formation by a process 
whose time-scale is long compared to the migration time-scale.

In the case of giant planet formation by the core accretion process
\citep[e.g.,][]{bodenheimer1986,wuchterl1991a}, the formation time-scale of 
about $10^7$ years \citep{pollack1996,tajima1997} is rather long compared 
to migration time-scales of about $10^5$ years 
\citep[e.g.,][]{lin1986,ward1997} 
for a Jupiter-mass planet and about $10^6$ years for an Earth-mass 
core (\citealp*{tanaka2002}; \citealp*{gennaro2003b}; \citealp{bate2003}).

However, it has been found recently \citep*{rice2003,alibert2004} that
the effects of accretion and migration of a planetary core
can significantly reduce the time needed by the core to reach the 
mass necessary for the nucleated instability to occur 
\citep{wuchterl1991b,magni2004}.
Furthermore,
several recent studies have suggested that additional effects may 
be of importance to migration. These include thermal effects of the
disc material near a planet 
\citep{morohoshi2003,jang-condell2004}, effects 
of radial opacity jumps in the disc \citep{menou2004}, effects of 
vortices induced by a planet \citep*{koller2003}, effects of turbulent
fluctuations \citep{rnelson2004}, and effects of coorbital
material \citep[][ hereafter MP03]{masset2003}.
In the current study, we consider effects of coorbital material, along
the lines of MP03.

Corotation torques arise in the coorbital region. In the absence of
dissipation or other time-dependent effects, the corotation torque is
zero in a smooth disc.  The reason is that in a steady-state, fluid
elements circulate in closed orbits. Over a libration time-scale, a
fluid element will gain and lose torque, but the result is zero
average torque. Formally, the corotation region in linear theory gives
rise to a torque that depends on the gradient of the disc vortensity
\citep[e.g.,][]{gt1979}. This torque is properly interpreted as an
``unsaturated'' or 
maximal torque that arises over time-scales less than a libration 
time-scale or when the effects of viscosity are sufficiently 
large in steady-state.
A derivation that includes nonlinear feedback shows that the steady-state 
corotation torque is indeed zero for a fluid in a smooth inviscid disc
\citep[ see also \citealp{masset2001}]{balmforth2001,ogilvie2003}.
But, even the unsaturated corotation torque for typical planet-disc
systems is somewhat smaller in magnitude than the other (Lindblad) 
torques present \citep{tanaka2002}.
Furthermore, for typical disc parameters, this torque is likely 
saturated (reduced to a smaller value), since the effects of turbulent
viscosity are not sufficiently strong, at least in an alpha model 
description.

The above-described analyses did not take into account the
effects of the radial migration of the planet.
This motion may cause enough asymmetry in the
corotational flow that a net torque occurs, which
may lead to a ``runaway'' situation (MP03). That is,
the migration of the planet might cause a corotational torque that
enhances the migration rate, which in turn further promotes asymmetry 
and leads to a stronger torque, etc. 
Examples of such a runaway phenomenon were reported in simulations
by MP03. The most favourable circumstances for such a process are
expected when a planet interacts with a massive disc in which there 
is not a clean gap.

In addition to the classical corotational torques that arise from
nearly librating orbits,
coorbital torques can also arise within the Hill sphere of the planet. 
Material flows into this region and forms a circumplanetary disc with 
shocks \citep*{lubow1999,gennaro2002}.

Our previous studies did not allow the planet to migrate during the
course of the simulation and therefore could not have found such a
runaway migration. Numerical resolution is a key issue because
densities near a planet are relatively high and fractionally small 
density errors there can give rise to large spurious torques.  
\citet{bate2003} found that torques near the planet may contribute 
somewhat ($\sim 20$ per cent) to the migration rate. However, that 
study lacked the resolution to reliably determine such torques.

In this paper, we investigate if the torques exerted on a
high-mass planet by a disc depend significantly on whether the planet
is kept on a fixed orbit or allowed to migrate.  We also
investigate the possible role of torques due to material within
the Hill sphere. We do this by means
of two-dimensional (2D) and three-dimensional (3D) high resolution
hydrodynamical simulations. A key feature of the code is that it
allows high resolution to be achieved by means of nested grids that
encompass a region around the planet as it migrates.  
With this code, we are able to examine the contribution of the
material inside the planet's Hill sphere to the total torque on the 
planet.

\pagestyle{myheadings}
\markboth{\hfill Migration rates of protoplanets \hfill}%
       {\hfill \textsc{G. D'Angelo, M. Bate, \& S. Lubow} \hfill}
The outline of the paper is as follows.
In \refSect{sec:physical_model} the physical model is described.
In \refSect{sec:numerical_method} we present an overview of the numerical 
procedures employed in these computations. The results of the calculations 
are provided in \refSects{sec:lm_models}{sec:hm_models}.
In \refSect{sec:discussion} we present a discussion of these results
and our conclusions.

\section[]{Description of the physical model}
\label{sec:physical_model}
It is generally believed that the interaction between a circumstellar disc 
and a Jupiter-sized object can be studied by means of a two-dimensional 
approximation (\citealp*{kley2001}; \citealp{gennaro2003b}).
However, while this is possibly true when considering interactions occurring 
at Lindblad resonance locations (i.e., at distances from the planet larger 
than a disc scale-height, $H$), it is not yet clear whether or to what extent 
this remains a valid assumption when dealing with other interactions occurring 
at coorbital locations \citep{masset2002}.
Therefore, in this investigation we considered both 2D and 3D disc models. 

In the 2D geometry we employed a cylindrical coordinate frame 
$\{O; r, \phi, z\}$, 
with the disc confined in the plane $z=0$, whereas in the 3D geometry we used 
a spherical polar coordinate frame  $\{O; R, \theta, \phi\}$.  The rotational 
axis of the disc is either parallel to the $z$-axis or to the polar direction, 
$\theta=0$. Both reference frames have their origin, $O$, on the star and 
rotate about the disc axis with an angular velocity $\gOmega$ and an angular
acceleration $\gdotOmega$, being this last vector also parallel to the disc 
axis. The magnitudes of $\gOmega$ and $\gdotOmega$ are specified later in this 
section. For the sake of clarity we point out that, whenever the variable $r$ 
is used in the context of spherical polar coordinates, it 
indicates the distance from the rotational axis $r=R\,\sin{\theta}$. 
\subsection[]{Equations of motion for the disc}
\label{sec:equations_of_motion_disk}
The hydrodynamical equations describing the disc evolution are usually
written in the conservative form for the radial and angular momenta.
This can be derived from the Navier-Stokes equations for the velocities
\citep[see, e.g.,][ Chapter~3]{m&m} and the continuity equation.
Since the 2D equations in \textit{cylindrical coordinates} can be formally
derived from the 3D equations in \textit{spherical polar coordinates},
we explicitly write them only for the latter reference frame.
Indicating with $\rho$ the mass density, 
with $\gvec{u}\equiv(u_{R},u_{\theta},u_{\phi})$ the fluid velocity,
and with $\omega_{\rmn{A}}=\omega+\Omega$ the \textit{absolute} angular
velocity of the fluid around the disc axis ($\omega\,r=u_{\phi}$), 
the equations of motion for the disc in conservative form can be written as
\begin{equation}
 \label{eq:cnt}
 \ddt{\rho} + \gdiv(\rho\,\gvec{u})=0,
\end{equation}
\begin {equation}
  \label{eq:xi_R}
  \ddt{\xi_{R}} + \gdiv(\xi_{R}\,\gvec{u})
   =  \rho\,(\frac{u^2_{\theta}}{R}
   +  \omega^2_{\rmn{A}}\,R\,\sin^2{\theta})
   -  \dd{p}{R} - \rho\,\dd{\Phi}{R} + f_{R},
\end{equation}
\begin{equation}
 \label{eq:xi_theta}
 \ddt{\xi_{\theta}} + \gdiv(\xi_{\theta}\,\gvec{u})
  =  \rho\,\omega^2_{\rmn{A}}\,R^2\,\frac{\sin{2\,\theta}}{2}
  -  \dd{p}{\theta} - \rho\,\dd{\Phi}{\theta}
  +  R\,f_{\theta},
\end{equation}
\begin {equation}
\label{eq:xi_phi}
 \ddt{\xi_{\phi}} + \gdiv(\xi_{\phi}\,\gvec{u})
  = - \dd{p}{\phi}- \rho\,\dd{\Phi}{\phi}
  + R\,\sin{\theta}\,f_{\phi},
\end{equation}
where 
\begin{equation}
  \label{eq:mv3d}  
  (\xi_{R},\xi_{\theta},\xi_{\phi})=%
              \rho\,(u_{R},%
                     u_{\theta}\,R,%
                     \omega_{\rmn{A}}\,R^2\,\sin^2{\theta})
\end{equation}
are the radial and angular momentum densities.
Equations in 2D cylindrical coordinates can be obtained from 
equations~(\ref{eq:cnt}), (\ref{eq:xi_R}), (\ref{eq:xi_phi}), and 
(\ref{eq:mv3d}) by replacing $\rho$ with the surface density $\Sigma$, 
using the appropriate expression for the divergence operator, dropping
all terms that contain the velocity $u_{\theta}$, and setting 
$\theta=\pi/2$.

Note that $\xi_{\phi}$ is the \textit{absolute} azimuthal angular momentum 
(density) of the fluid rather than that relative to the rotating reference 
frame. 
This basically means that the non-inertial terms arising from the
rotation of the reference frame (i.e., Coriolis and angular velocity 
accelerations) are incorporated in the left-hand side of \refeqt{eq:xi_phi}.
This choice assures a better numerical treatment of the associated 
conservation law \citep{kley1998}.

We adopted a locally isothermal equation of state by setting
$p=c^2_{\rmn{s}}\,\rho$ (or $p=c^2_{\rmn{s}}\,\Sigma$ in 2D) . 
The sound speed, $c_{\rmn{s}}$, is equal to the disc aspect ratio, 
$H/r$, times the Keplerian velocity, $v_{\rmn{K}}$. We used a constant 
disc aspect ratio throughout the disc, implying that the temperature 
distribution scales as the inverse of the distance from the disc axis.

Since self-gravity is ignored, the gravitational potential,$\Phi$, only 
includes contributions from the star, the planet, and the non-inertial 
forces arising from the motion of the frame origin, $O$.
Indicating the position vector of a fluid element as $\gvec{x}$ and that 
of the planet as $\gvec{x}_{\rmn{p}}$, the disc gravitational potential 
reads
 \begin{equation}
   \label{eq:phi}
   \Phi=%
        - \frac{G\,\MStar}{|\gvec{x}|}%
        - \frac{G\,\Mp}{\sqrt{\mdp^2+\varepsilon^2}}%
        + \frac{G\,\Mp}{|\gvec{x}_{\rmn{p}}|^3}%
        \,\gvec{x}\bmath{\cdot}\gvec{x}_{\rmn{p}},
 \end{equation}
where $\MStar$ is the stellar mass, $\Mp$ is the planet mass, and 
$\varepsilon$ is a smoothing length (see the discussion in
\refSecp{sec:physical_par}).
The third term on the right-hand side of \refeqt{eq:phi} originates 
from the fact that the origin of the coordinate frame is accelerated 
by the planet\footnote{%
To be strict, an additional term should appear in \refeqt{eq:phi} due 
to the force exerted by the disc material on to the star, as measured 
from the centre-of-mass reference frame. We neglected this contribution, 
as is done when assuming that the centre-of-mass of the whole system
coincides with that of the star--planet system.}.

The viscosity force density, $\gvec{f}\equiv(f_{R},f_{\theta},f_{\phi})$
(or $\gvec{f}\equiv(f_{r},f_{\phi})$ in 2D), is
written as $\gvec{f}=\nabla\bmath{\cdot}\mathbfss{S}$. It assumes a 
standard viscous stress tensor, $\mathbfss{S}$, for a Newtonian fluid 
with a constant kinematic viscosity, $\nu$, and a zero bulk viscosity. 
Explicit forms for the components of $\gvec{f}$ can be found in 
\citet[ Chapter~3]{m&m}, for the 3D spherical polar coordinates case 
and in \citet{gennaro2002}, for the 2D cylindrical coordinates case.
\subsection[]{Equation of motion for the planet}
\label{sec:equation_of_motion_planet}
In the present study the planet's orbit evolves under the gravitational 
action of the central star and of the disc material. Moreover, since the 
orbit is described with respect to a varying rotating reference frame, 
all non-inertial terms involving the angular velocity, $\gOmega$, and the 
angular acceleration, $\gdotOmega$, of the coordinate system have to be 
taken into account. Restricting to those orbits 
\textit{coplanar with the disc midplane} ($\theta=\pi/2$ or $z=0$),  the 
equation of motion of the planet is
\begin{eqnarray}
  \label{eq:pme}
  \ddot{\gvec{x}}_{\rmn{p}} &=&
      -\frac{G(\MStar+\Mp)}{|\gvec{x}_{\rmn{p}}|^3}\,\gvec{x}_{\rmn{p}}%
       +\gvec{\mathcal{A}}_{\rmn{p}}-\gvec{\mathcal{A}}_{*}\\
                            & &
       -\gOmega\bmath{\times}%
        \left(\gOmega\bmath{\times}\gvec{x}_{\rmn{p}}\right)%
       -2\,\gOmega\bmath{\times}\dot{\gvec{x}}_{\rmn{p}}%
       -\gdotOmega\bmath{\times}\gvec{x}_{\rmn{p}}.\nonumber
\end{eqnarray}
We recall that, by working hypothesis, $\gOmega$ as well as $\gdotOmega$ 
are perpendicular to the disc midplane and produce a counter-clockwise 
rotation. The acceleration applied by the disc matter to the planet is 
given by
\begin{equation}
  \label{eq:A2}
  \gvec{\mathcal{A}}_{\rmn{p}}=%
   G\int_{\Md} \frac{\left(\vdp\right)\,\rmn{d}\Md(\gvec{x})}%
                    {\left(\mdp^2 + \varepsilon^2\right)^{3/2}}, 
\end{equation}
while the acceleration applied to the star is
\begin{equation}
  \label{eq:A1}
  \gvec{\mathcal{A}}_{*}=%
   G\int_{\Md} \frac{\gvec{x}\,\rmn{d}\Md(\gvec{x})}%
                    {|\gvec{x}|^{3}}.
\end{equation}
In both cases the integration is carried out over the simulated
disc mass, $\Md$ (see \refSecp{sec:physical_par}).

\refEqt{eq:A2} contains the smoothing length in its denominator.
This expression for acceleration appears in the second term of 
\refeqt{eq:phi}. An acceleration with smoothing is applied to the 
planet's motion in order to satisfy Newton's third law.

\subsection[]{Rotational elements of the reference frame}
\label{sec:rot_elements}
The main aim of this paper is to study the exchange of angular momentum 
occurring between a migrating planet and disc material moving on U-turns 
of horse-shoe orbits. In order to accurately resolve the flow variables 
in this region by means of a local grid-refinement technique, the planet 
needs to move through the grid as slowly as possible. To achieve this, we 
worked in reference frames that rotate about the disc axis at a variable rate, 
$\Omega=\Omega(t)$. We then chose $\Omega$ and $\dot{\Omega}$ so as to 
compensate for the fastest component of the planet motion, i.e., the 
azimuthal one. This is accomplished by calculating the \textit{total} 
orbital angular momentum of the planet per unit mass, 
$\gvec{\mathcal{H}}_{\rmn{A}}$, and then requiring that
\begin{eqnarray}
  \label{eq:rotelOm}
  \gvec{\mathcal{H}}_{\rmn{A}}       & =&%
                                   \gvec{x}_{\rmn{p}}\bmath{\times}%
                                  (\gOmega\bmath{\times}\gvec{x}_{\rmn{p}}),\\
  \label{eq:rotelOmdot}
  \dot{\gvec{\mathcal{H}}}_{\rmn{A}} & = &%
                                   \gvec{x}_{\rmn{p}}\bmath{\times}%
                                  (\gvec{\mathcal{A}}_{\rmn{p}}%
                                  -\gvec{\mathcal{A}}_{*}).
\end{eqnarray}
These equations are to be solved with the additional requirements
that both $\gOmega$ and $\gdotOmega$ must be perpendicular to the 
plane of the orbit and produce a positive (i.e., counter-clockwise) 
rotation.
Equations~\ref{eq:rotelOm} and \ref{eq:rotelOmdot} constrain the
angular velocity and acceleration of the rotating coordinate system 
so that the planet trajectory reduces to a purely radial motion.
In other terms, all of the planet's orbital angular momentum is 
conveyed to the rotation of the non-inertial reference frame.
 
If the orbital eccentricity remains close to zero during the system
evolution, as we found in our simulations, then the planet's radial motion 
is only due to the disc gravitational torques. We denote the planet's 
semi-major axis  as $a=\mxp$ and the time-scale of this drifting motion 
as $\taum=a/|\dot{a}|$. 
The quantity $N_r\,\Delta r$ (or $N_R\,\Delta R$) is the radial extent 
of the highest refinement region and the time spent within this region 
is $N_r\,\Delta r/|\dot{a}|=N_r\,(\Delta r/a)\,\taum$, which is on the 
order of $0.1\,\taum$ for the parameters used in the calculations.
Numerical simulations 
\citep[e.g.,][]{lubow1999,rnelson2000,kley2001,gennaro2002} 
as well as analytical theories \citep{gt1980,lin1986,ward1997} on 
disc torques suggest time-scales, $\taum$, on the order of $10^4$ 
periods.
Therefore, with this method one can expect to track the planet and 
the coorbital regions, with the necessary numerical resolution, for
hundreds of orbits.
\subsection[]{Physical parameters}
\label{sec:physical_par}
We performed two kinds of simulations: the first kind is dedicated to 
planets interacting with a low-mass disc and the second is dedicated
to planets orbiting in a high-mass disc.
In all of the calculations, the mass of the star, $\MStar$, represents 
the unit of mass whereas the initial semi-major axis of the planet's orbit, 
$a_{0}$, gives the length unit. 
The unit of time is such that $1/t_{0}=\sqrt{G\,(\MStar+\Mp)/a^{3}_{0}}$.
However, when it is necessary to convert quantities into physical units, 
we used $\MStar=1\,\MSun$ and $a_{0}=5.2\,\AU$.
\subsubsection[]{Parameters for low-mass discs}
\label{sec:lm_par}
In these models the simulated disc domain extends radially from $0.4$ to 
$4.0$ length units around the star and, azimuthally in angle, from $0$ to 
$2\pi$. 
These simulations describe a disc of mass $\Md=7.5\times10^{-3}\,\MStar$  
within the radial limits of the computational domain, which
is equivalent to $0.01\,\MSun$ within $26\,\AU$ of a
$1\,\MSun$ star.
In the case of 3D models, we simulated only the upper half of the disc between
$80^{\circ}\le\theta\le90^{\circ}$ and assumed  mirror symmetry with
respect to the midplane. The aspect ratio of the disc was fixed to $H/r=0.05$.
The overall initial surface density scales as $r^{-1/2}$ and is axisymmetric.
This would give an unperturbed disc surface density at the location of the
planet of $76\,\sdunits$,
but we included an initial gap along the planetary orbit that accounts for 
an approximate balance of viscous and tidal torques.
One model was also run without an initial gap, in order to determine
its influence on the results.
In 3D models, the initial latitude dependence of the mass density is taken 
to be a Gaussian.

We employed a constant kinematic viscosity, $\nu$, to account for the effects 
due to turbulence in the disc. In the units introduced above, we set 
$\nu=10^{-5}$ that is also equivalent to Shakura \& Sunyaev parameter 
$\alpha=4\times10^{-3}$ at the initial location of the planet. This choice is 
compatible with what was recently found in studies of embedded Jupiter-size 
bodies in discs with MHD turbulence \citep{papa2003,winters2003}. However,
we do not include the spatial variations in $\alpha$ consistent with the MHD 
results, nor the time fluctuations due to MHD turbulence \citep{rnelson2004}.

The planet mass is such that $\Mp/\MStar=10^{-3}$ (i.e., one Jupiter-mass, 
$\MJup$, for a one-solar-mass star). The planet starts on a circular orbit of 
semi-major axis $a_{0}=1$, which is kept static for a certain number of periods 
to allow the relaxation of the system. This was done by setting to zero the 
terms (\ref{eq:A2}) and (\ref{eq:A1}), in \refeqt{eq:pme}, and activating them 
at the ``release'' time, $t=t_{\rmn{rls}}$.  We used $t_\rmn{rls}$ equal to 
either $100$ or $300$ orbits. 
The migration rates were found to be insensitive to the release 
time (less than $10$ per cent differences in rates), provided it is greater
than $100$ orbits. The azimuthal position of the planet remains constant 
throughout the computations (see \refSecp{sec:rot_elements}) and it is equal 
to $\phi=\phi_{\rmn{p}}=\pi$.

The smoothing length of the planet potential, $\varepsilon$, in \refeqt{eq:phi} 
was chosen to be a fraction of the planet's Hill radius, 
$\Rhill=a\,\left[\Mp/(3\,\MStar)\right]^{1/3}=0.069\,a$. We employed three 
different values: $\varepsilon=0.4$, $0.2$, and $0.1\,\Rhill$, in order to
study the effects of smoothing on the results.
\subsubsection[]{Parameters for high-mass discs}
\label{sec:hm_par}
When simulating planets embedded in a high-mass disc, we used parameters as 
similar as possible to those adopted by MP03, in order to have a direct 
comparison with their results. Therefore, in contrast to the previous settings, 
the radial extent of the disc and its aspect ratio were reduced to $[0.4,2.5]$ 
length units and $0.03$, respectively.
The simulations describe a disc of mass
$\Md=2.37\times10^{-2}\,\MStar$ within the radial limits of the
computational domain, which
is equivalent to $\approx24\,\MJup$ within $13\,\AU$
of a $1\,\MSun$ star. As in MP03, the initial
surface density scales as $r^{-3/2}$ and there is no initial gap. This gives 
an unperturbed disc surface density at the location of the planet of 
$653\,\sdunits$. The planet-to-star mass ratio is $\Mp/\MStar=3\times10^{-4}$, 
roughly corresponding to a Saturn-mass object for $\MStar=1\,\MSun$. We again 
employed a constant kinematic viscosity $\nu=10^{-5}$ in dimensionless units.
The planet was held on a static orbit ($a_{0}=1$) and released at 
$t=t_{\rmn{rls}}$. For most of the 2D calculations the planet was released 
after $477$ orbits, as done by MP03. For comparisons between 2D and 3D models
we could not afford the time required to run 3D calculations to $477$ orbits, 
so we released the planet at $200$ orbits.
For a convergence test with high-resolution 2D calculations we set 
$t_{\rmn{rls}}=277$ orbits.
The value of the smoothing length was set to 
$\varepsilon=0.3878\,\Rhill$ ($\Rhill=0.046\,a$), which is equal 
to $60$ per cent of the local disc scale-height. 
\section[]{Description of the numerical method}
\label{sec:numerical_method}
The hydrodynamical equations~(\ref{eq:cnt}) through (\ref{eq:mv3d}) that 
describe the evolution of the 
disc are solved numerically by means of a finite-difference scheme with 
directional operator splitting. The method is second-order accurate in space 
and first-order in time \citep{ziegler1997}. The numerical resolution of the 
regions around the planet is greatly enhanced by utilising a nested-grid 
technique \citep[for details, see][]{gennaro2002,gennaro2003b}. Each subgrid 
level increases the resolution, with respect to the hosting grid, by a factor 
$2$ in each direction.  Thus, the total gain in resolution for each added
subgrid is $2^2$ or $2^3$ in 2D or 3D simulations, respectively.
Subgrids are \textit{fully} nested, i.e., each occupies a region of space 
completely contained inside the hosting grid. This implies that the number 
of zones of any subgrid, along any direction, can be at most twice the number 
of zones of the hosting grid along that direction. A point in space is handled 
by the highest resolution grid (highest grid level) that covers that point.

In order to test the behaviour of the nested-grid code for
planetary
migration calculations, we compared outcomes of models executed in a 
single-grid mode with those of the same models executed in a nested-grid
mode with equal numerical resolution. We always found an excellent agreement
with discrepancies averaging $\approx 10^{-3}$ per cent. Some of these
comparisons are reported in the Appendix.

The equations of motion of the planet are solved in Cartesian coordinates 
with a high-accuracy and fast hybrid algorithm. This involves a Bulirsch-Stoer 
method with an adaptive time-step control \citep{press1992} and a standard 
fourth-order Runge-Kutta solver. Each hydrodynamics time-step $\Delta t$ 
(constrained by the Courant-Friedrichs-Lewy stability criterion) is divided 
into substeps whose duration is dictated by the requirement that the local
truncation error is always smaller than the chosen accuracy ($10^{-7}$ in 
these calculations). The maximum number of substeps allowed is set to $5000$.
If the integration time has not reached the value $\Delta t$ after this 
iteration cycle, the remainder of the time-interval is integrated via a
fourth-order Runge-Kutta method. 
Although this is a necessary precaution, the overall procedure actually 
requires only a few time-substeps of the Bulirsch-Stoer algorithm to complete 
the whole hydrodynamics time-step $\Delta t$, since the vector \refeqt{eq:pme} 
has no singular points inside this integration interval.
The orbit integrator was tested, over long-term evolutions, against both 
circular and eccentric Keplerian orbits. For a variety of values
of $\Omega$ and $\dot{\Omega}$, no deviations from the analytic solutions 
were found down to the machine precision.

Disc gravitational forces given by \refeqs{eq:A2}{eq:A1} are considered to be 
constant over the whole time span $\Delta t$ and are computed by summation of 
discretised quantities over the whole grid, always using densities from the 
subgrid with the highest resolution available.
\subsection[]{Numerical setup}
\label{sec:numerical_setup}
\begin{table}[!t]
 \begin{center}
 \caption{Grid system employed in low-mass disc models.}
 \label{tbl:grids1}
                    \resizebox{1.0\linewidth}{!}{%
 \begin{tabular}{@{}ccccc@{}}
  \hline
  Grid  & 2D3G                     & 2D4G                     & 2D5G                  &
          3D3G \\
  level & $N_{r}\times N_{\phi}$   & $N_{r}\times N_{\phi}$   & $N_{r}\times N_{\phi}$&
          $N_{R}\times N_{\theta}\times N_{\phi}$\\
  \hline
   $1$  & $243\times 455$          & $243\times 455$          & $243\times 455$       &
          $243\times 17\times 455$ \\
   $2$  & $114\times 84\phantom{0}$& $114\times 84\phantom{0}$& $114\times 104$       &
          $114\times 24\times 84\phantom{0}$ \\
   $3$  & $114\times 84\phantom{0}$& $114\times 84\phantom{0}$& $134\times 124$       &
          $114\times 24\times 84\phantom{0}$ \\
   $4$  &                          & $134\times 84\phantom{0}$& $154\times 144$       &
                                  \\
   $5$  &                          &                          & $174\times 164$       &
                                  \\
  \hline
 \end{tabular}
                                                }
 \end{center}
\small{%
 The linear resolution around the planet, averaged over all directions, 
 on the level $1$ is $1.45\times10^{-2}$. This value decreases by a factor of
 $2^{(l-1)}$ on a given level $l$. Thus,
 the grid systems 2D3G and 3D3G resolve the flow around the Hill sphere of a
 Jupiter-mass planet with $19$ grid zones per Hill radius, while the grid  
 2D5G achieves a resolution of $76$ zones per Hill radius.
}
\end{table}
\begin{table*}
 \begin{center}
 \begin{minipage}{0.75\linewidth}
 \caption{Grid system utilised in high-mass disc models.}
  \medskip
 \label{tbl:grids2}
                    \resizebox{\linewidth}{!}{%
 \begin{tabular}{@{}ccccccc@{}}
  \hline
  Grid  & 2D1Gb                    & 2D3Gb                    
        & 2D4Gb                    & 2D5Gb                    
        & 2D6Gb                    & 3D3Gb \\
  level & $N_{r}\times N_{\phi}$   & $N_{r}\times N_{\phi}$   
        & $N_{r}\times N_{\phi}$   & $N_{r}\times N_{\phi}$   
        & $N_{r}\times N_{\phi}$   & $N_{R}\times N_{\theta}\times N_{\phi}$\\
  \hline
   $1$  & $147\times 455$          & $147\times 455$          
        & $147\times 455$          & $147\times 455$          
        & $147\times 455$          & $147\times 17\times 455$ \\
   $2$  &                          & $114\times 84\phantom{0}$
        & $114\times 84\phantom{0}$& $114\times 84\phantom{0}$
        & $134\times 104$          & $\phantom{0}84\times 24\times 84\phantom{0}$ \\
   $3$  &                          & $114\times 84\phantom{0}$
        & $114\times 84\phantom{0}$& $114\times 84\phantom{0}$
        & $134\times 104$          & $\phantom{0}84\times 24\times 84\phantom{0}$ \\
   $4$  &                          & 
        & $114\times 84\phantom{0}$& $134\times 84\phantom{0}$ 
        & $164\times 104$          & \\
   $5$  &                          &                          
        &                          & $164\times 104$          
        &  $194\times 124$         & \\
   $6$  &                          &                          
        &                          &           
        &  $324\times 204$         & \\
  \hline
 \end{tabular}
                                                }\\
  \medskip

\small{%
 The average linear resolution around the planet on the level $l$ is 
 $1.43\times10^{-2}/2^{(l-1)}$.
 Hence, the grid systems 2D3Gb and 3D3Gb resolve the flow around the Roche 
 lobe of a $0.3\,\MJup$ planet with $13$ grid zones per Hill radius while 
 the grid 2D6Gb achieves a resolution of more than $100$ zones per Hill radius.
 We used the single-level grid system, 2D1Gb, only 
 for purposes of comparison with the calculations reported in MP03, 
       }
 \end{minipage}
 \end{center}
\end{table*}
In a disc-planet interaction calculation, the largest spatial gradients
of the flow variables develop around and inside the planet's Hill sphere.
Over a distance of two Hill radii, the density can change by three or more
orders of magnitude \citep{gennaro2003b,bate2003} and the velocity field 
describes highly complex patterns \citep[e.g.,][]{lubow1999,tanigawa2002}.
In order to ascertain to what extent our numerical experiments depend
upon the numerical resolution in this region, we performed a convergence 
study in all cases.
To this aim, we set up a number of grid systems whose resolution in the
coorbital regions ranges from $19$ (or $13$, when 
$\Mp/\MStar=3\times10^{-4}$) to nearly $76$ (or $104$) grid zones per Hill 
radius (see details in Tables~\ref{tbl:grids1} and \ref{tbl:grids2}).

Most calculations were performed without allowing the planet to accrete.
When accretion is permitted, mass is removed from around the planet within 
a tenth of its Hill radius. The removal of mass occurs on a time-scale on 
the order of a tenth of the orbital period.

Boundary conditions at the inner radial border allow flow towards
the central star, as naturally happens in viscous accretion discs. The outer 
radial border is closed so that no inflow or outflow of material is permitted. 
At both radial edges of the disc, the flow is assumed to be Keplerian around 
the central star. This circumstance may occasionally lead to spurious, 
small-amplitude wave excitation at the outer edge of the disc, since material 
there has the tendency to orbit about the centre-of-mass of the system rather 
than around the central star \citep[see discussion in][]{rnelson2000}. 
However, the effects of such waves are not relevant since they do not propagate
for a significant distance from the disc edge. In 3D models, reflective and 
symmetric boundary conditions are applied at the highest latitudes and at the 
midplane, respectively. The velocity field in the disc is initialised with a 
Keplerian circulation, corrected for the grid rotation.  

It was pointed out by \citet{anelson2003a} that when the planet moves through
the grid cells, the smoothing length in \refeqt{eq:phi} needs to be larger 
than half the linear dimension of the grid zone. This is required in order 
to avoid unphysical effects on the planet's trajectory due to close encounters 
with grid centres.
In these calculations we used initial values of $\varepsilon$ that are at 
least $1.8$ times the average linear size of the grid zone from which 
the planet is released.
Furthermore, the series of convergence tests that we performed indicate
that the ratios between $\varepsilon$ and the average grid zone size are 
large enough not to affect the outcome of the simulations 
(see \refSecp{sec:lm_models}).

\section[]{Results of low-mass disc models}
\label{sec:lm_models}
Previous studies of migrating Jupiter-mass planets showed that the evolution 
of the orbital semi-major axis, $a=a(t)$, is dependent upon the resolution 
with which hydrodynamics variables are discretised on the computing mesh 
\citep{rnelson2000,anelson2003a}. As emphasised by \citet{anelson2003b},
this issue becomes even more important when torques in the coorbital
region are resolved, i.e.,
when the planetary gravitational smoothing lengths are a small fraction of
$\Rhill$ (see \refeqp{eq:phi}). 
The dependence of gravitational torques on the
numerical resolution is crucial to assess the reliability
of the outcomes. Therefore, we tackled this problem with a number of dedicated 
simulations, before investigating any possible physical effects of coorbital 
torques on the migration of giant planets.
\subsection[]{A convergence study}
\label{sec:lm_cs}
\begin{figure}
\centering%
\resizebox{1.0\linewidth}{!}{%
\includegraphics{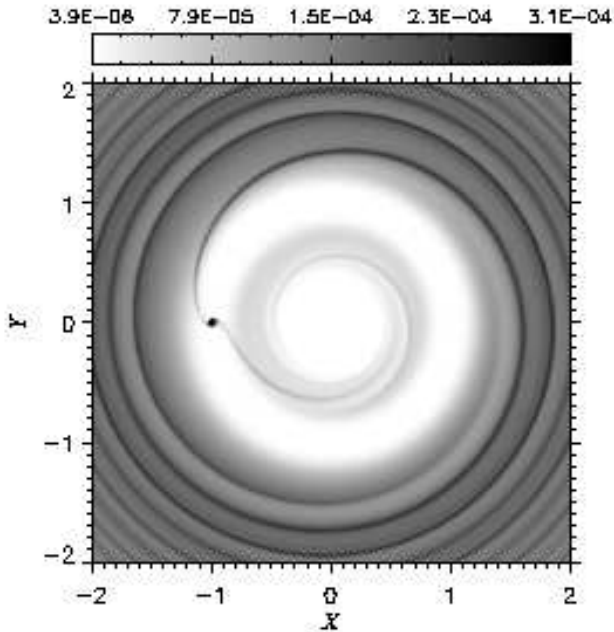}}
\caption{\small%
         Global surface density around a $\Mp=1\,\MJup$ planet
         orbiting in a low-mass disk  (see \refSecp{sec:lm_par}).
         The density is shown after $370$ orbits, when the planet
         has migrated for $70$ orbits. In the linear grey-scale,
         at $5.2\,\AU$, $\Sigma=10^{-4}$ corresponds to 
         $32.9\,\sdunits$. 
        }
\label{fig:jup_global}
\end{figure}
\begin{figure*}[t!]
\centering%
\resizebox{0.90\linewidth}{!}{%
\includegraphics{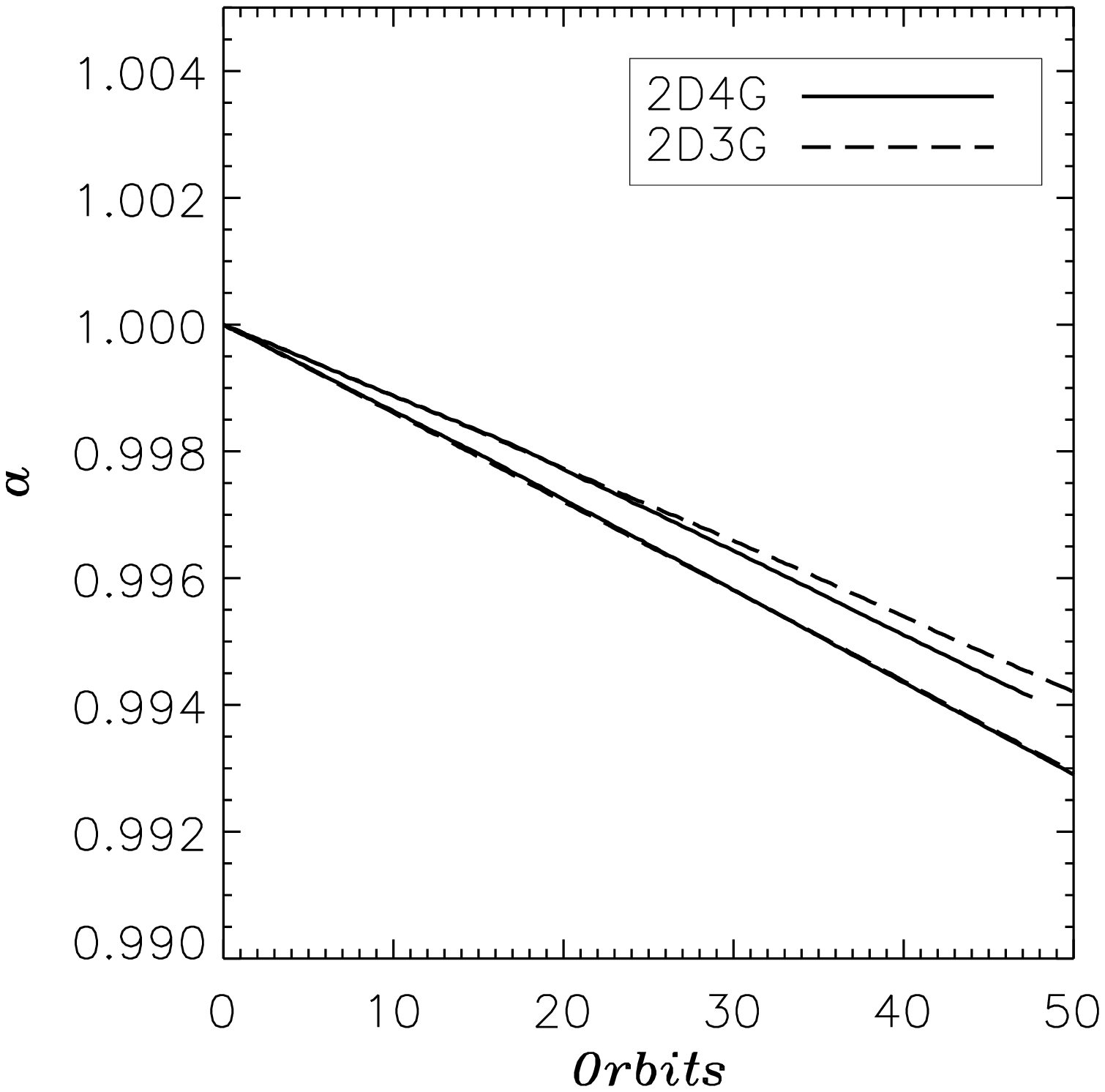}%
\includegraphics{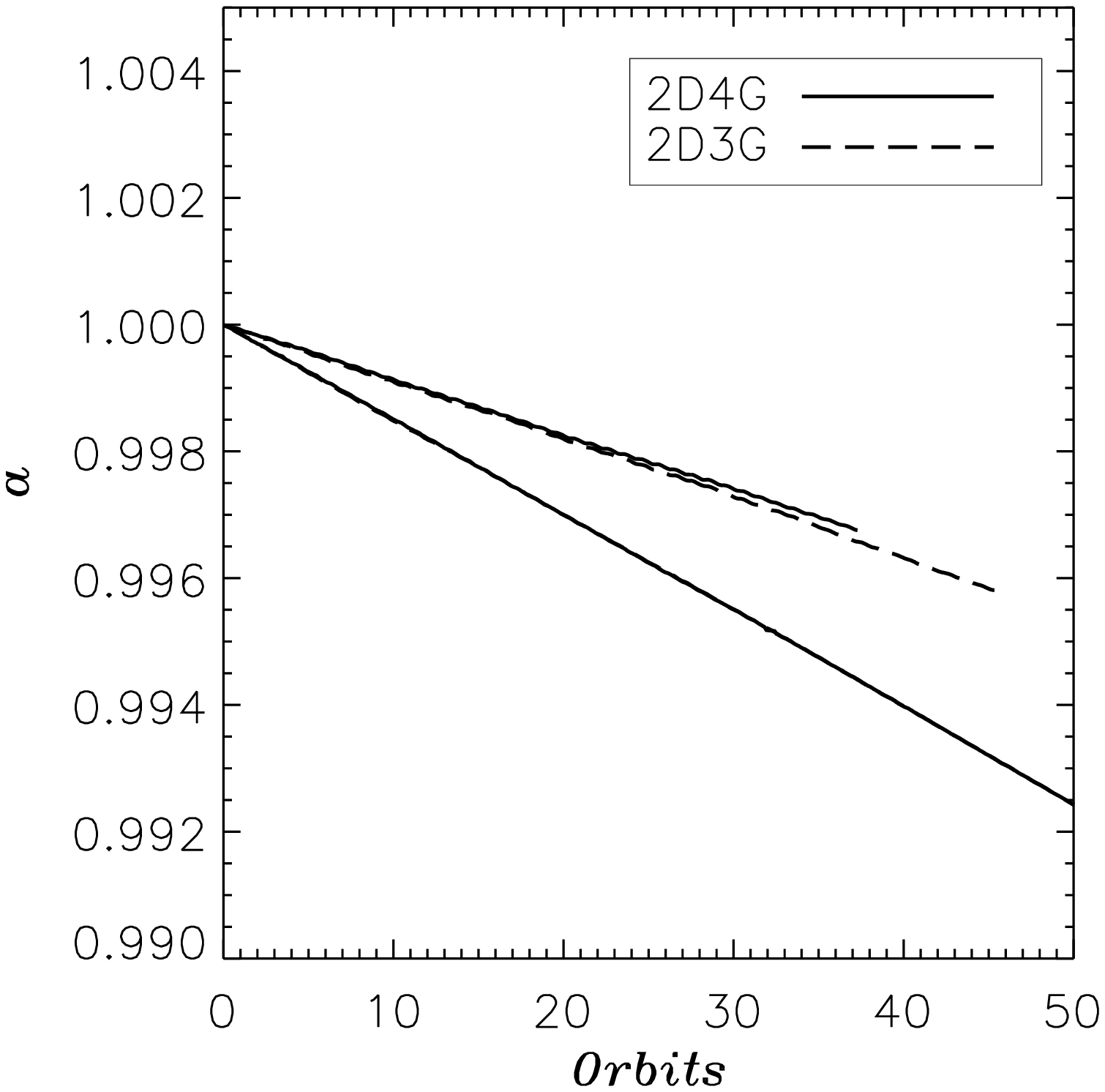}}
\resizebox{0.90\linewidth}{!}{%
\includegraphics{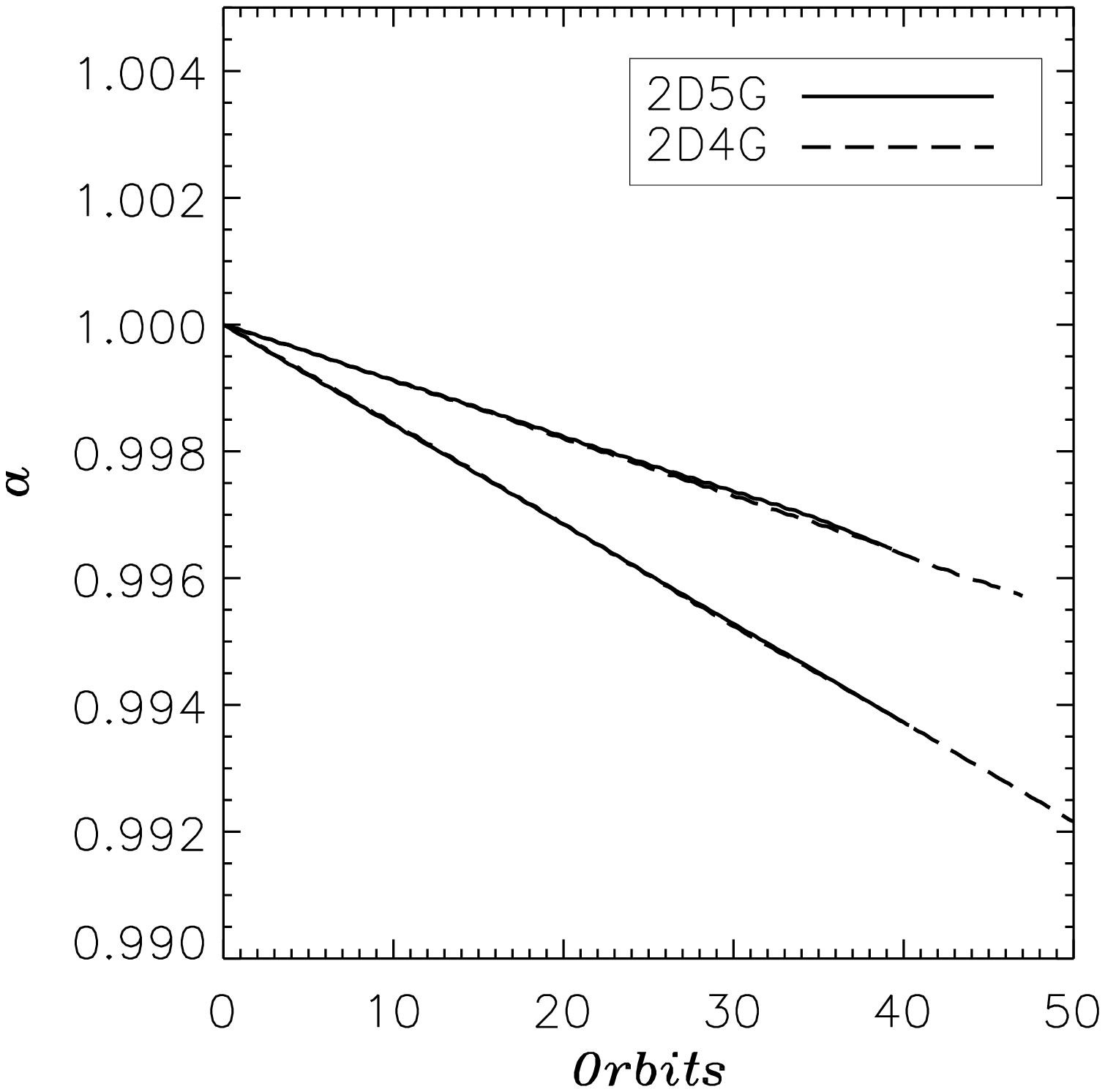}%
\includegraphics{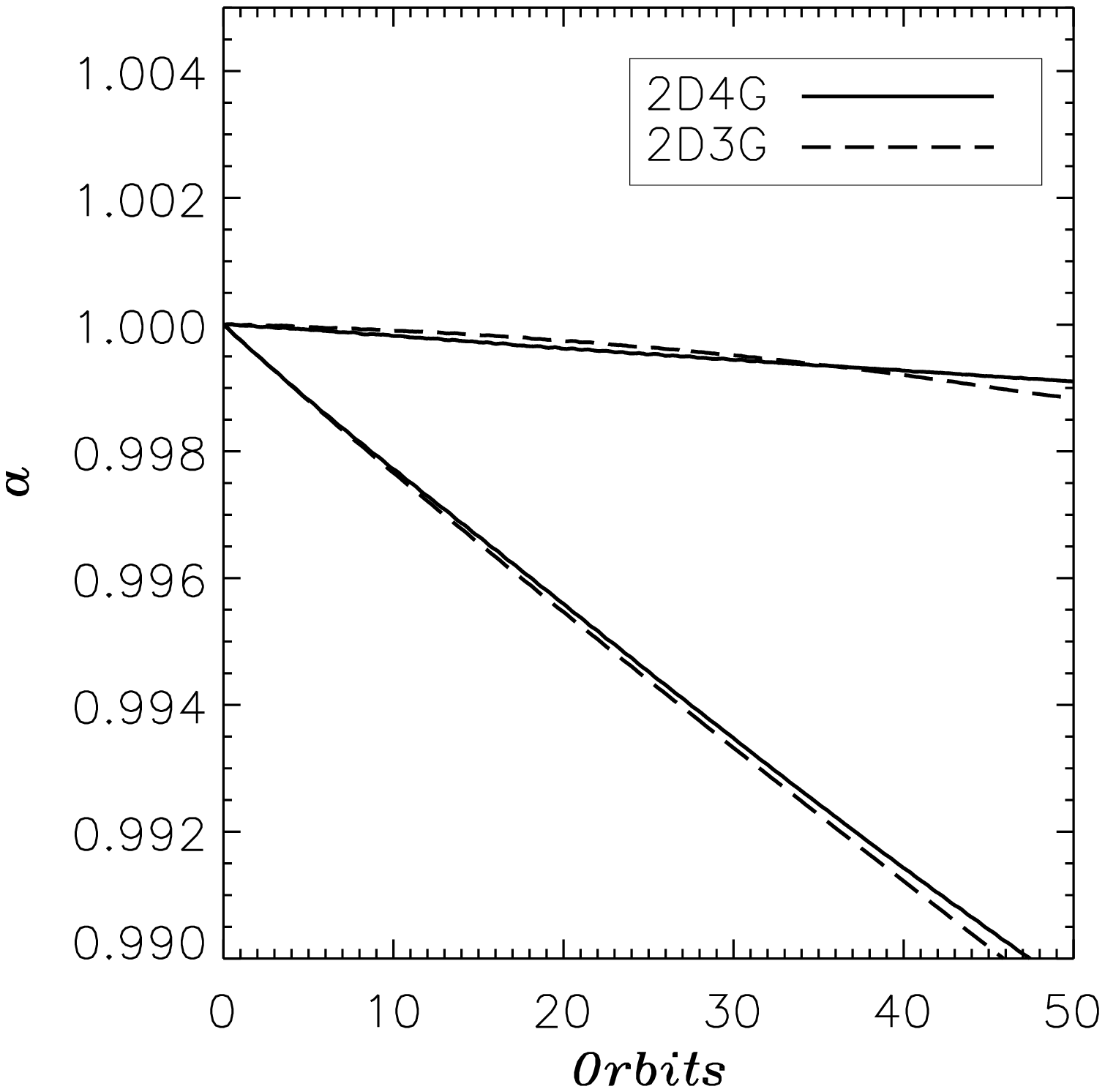}}
\caption{\small%
         Convergence tests regarding different configurations of Jupiter-mass 
         models orbiting in a two-dimensional low-mass disc.
         \textit{Upper-left}. Non-accreting planet with gravitational potential 
                               softening $\varepsilon=0.4\,\Rhill$.
         \textit{Upper-right}. Non-accreting planet with $\varepsilon=0.2\,\Rhill$. 
         \textit{Lower-left}. Non-accreting planet with $\varepsilon=0.1\,\Rhill$. 
         \textit{Lower-right}. Accreting planet with $\varepsilon=0.1\,\Rhill$.
         The release time is equal to $100$ orbits, except for the accreting
         model for which $t_{\rmn{rls}}=300$ orbits.
         Each panel shows how the semi-major axis, $a$, evolves
         when all torques from within the planet's Hill sphere are taken
         into account (upper curves) and when the contribution of those
         torques arising inside of $0.5\,\Rhill$ from the planet are
         neglected (lower curves). See text for further details.}
\label{fig:jup_acomp}
\end{figure*}
Convergence tests were carried out on each of the low-mass disc models 
described in \refSect{sec:lm_par}. The resolution was progressively increased 
by employing the grid systems 2D3G, 2D4G, and 2D5G (see \refTab{tbl:grids1}). 
The last two grid systems, compared to the first, provide
a linear resolution gain of a factor $2$ and $4$, respectively. 
For comparison purposes, we set the release time to $t_{\rmn{rls}}=100$ 
orbits, except for the simulations concerning the accreting model that have 
$t_{\rmn{rls}}=300$ orbits. 
The overall surface density from one of such calculations is displayed
in \refFgt{fig:jup_global}.

\begin{figure}
\centering%
\resizebox{1.0\linewidth}{!}{%
\includegraphics{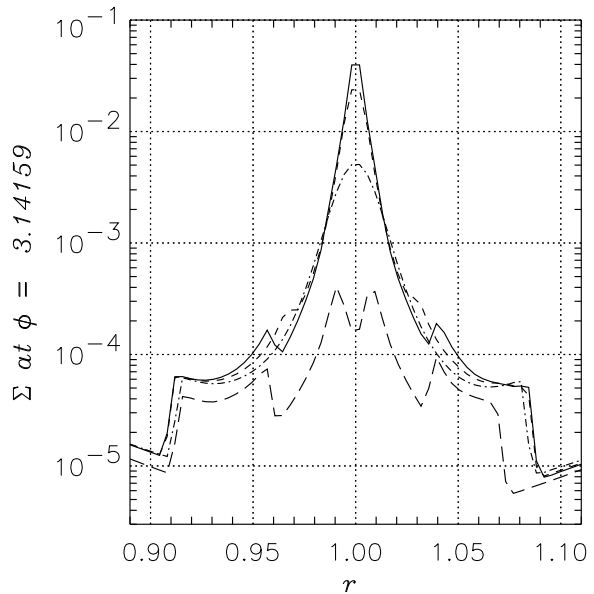}}
\caption{\small%
         Surface density profile along the azimuthal position 
         ($\phi=\phi_{\rmn{p}}$) of a $\Mp=1\,\MJup$ planet,
         after $100$ orbital periods, orbiting in a low-mass disc.
         Different line types indicate models with different values of
         the gravitational potential softening: 
         $\varepsilon=0.1\,\Rhill$ (\textit{solid line});
         $\varepsilon=0.2\,\Rhill$ (\textit{short-dash line});
         $\varepsilon=0.4\,\Rhill$ (\textit{dash-dot line});
         accreting planet (\textit{long-dash line}).
         If $a_{0}=5.2\,\AU$ and $\MStar=1\,\MSun$, $\Sigma=10^{-4}$ 
         corresponds to $32.9\,\sdunits$. 
        }
\label{fig:jup_dengra}
\end{figure}
\refFgt{fig:jup_acomp} shows that we achieved numerical convergence
in all cases, with either accreting or non-accreting planets and with 
different values of $\varepsilon$.
In some panels, the outcomes produced by the two grid systems can be hardly
distinguished.
The main numerical difficulty with the evaluation of gravitational torques 
arising from the coorbital region is related to the presence of 
large density gradients \citep{anelson2003b}. 
Moreover, the shorter the smoothing length, the larger such gradients are.
\refFgt{fig:jup_dengra} shows that there is an order-of-magnitude difference
between the density peaks of the models with $\varepsilon=0.1\,\Rhill$ and
$\varepsilon=0.4\,\Rhill$.  For the $\varepsilon=0.1\,\Rhill$ case, using
the grid system  2D3G would mean that $\varepsilon$ was resolved by less
than $2$ grid zones and, thus, the density gradients could be resolved too 
poorly. Therefore, we used the grid systems 2D4G and 2D5G for this case.

We also investigated whether there are any differences between simulations
starting with or without an initial density gap (see 
\refSecp{sec:physical_par}). Provided that the system is allowed to evolve
for a sufficiently long time in order that the gap becomes deep enough
($\approx 500$ orbits), the migration behaviour is very similar to that
of models initiated with a density gap.

\begin{figure*}[t!]
\centering%
\resizebox{0.90\linewidth}{!}{%
\includegraphics{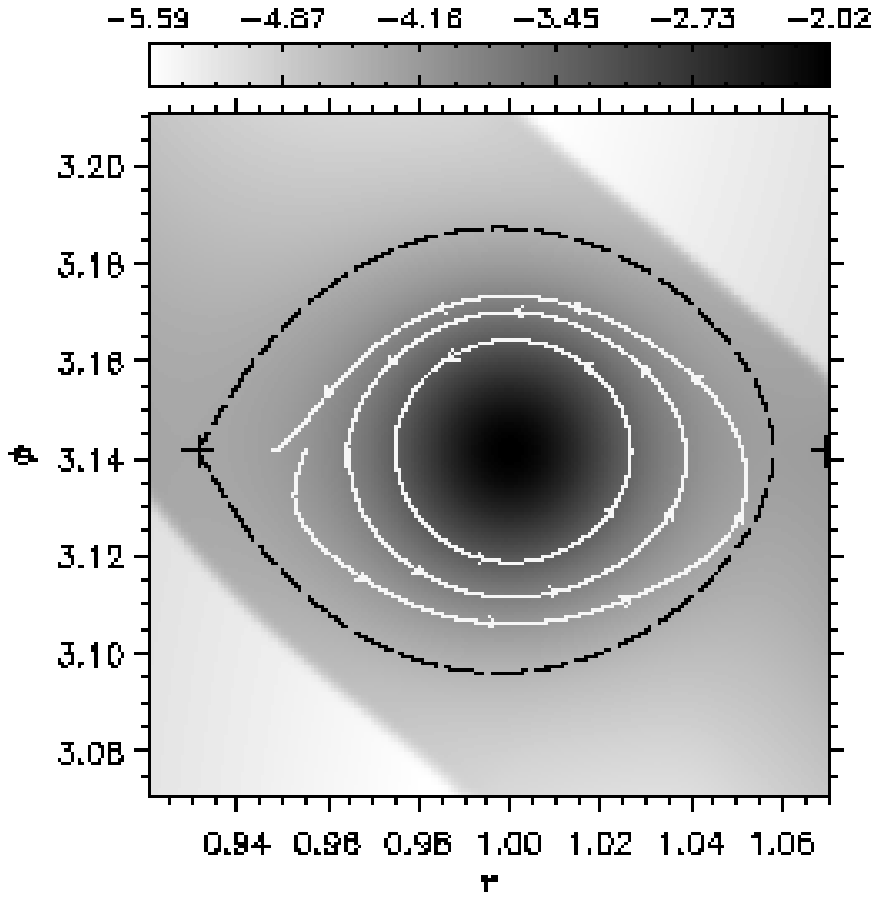}%
\includegraphics{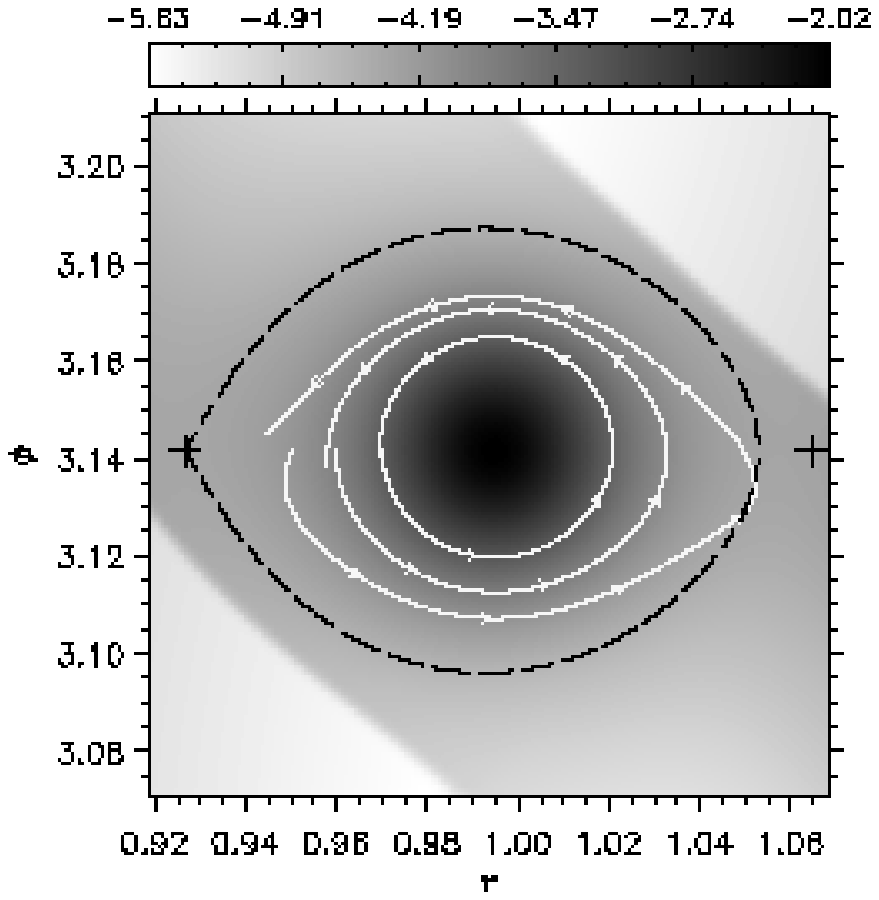}}
\resizebox{0.90\linewidth}{!}{%
\includegraphics{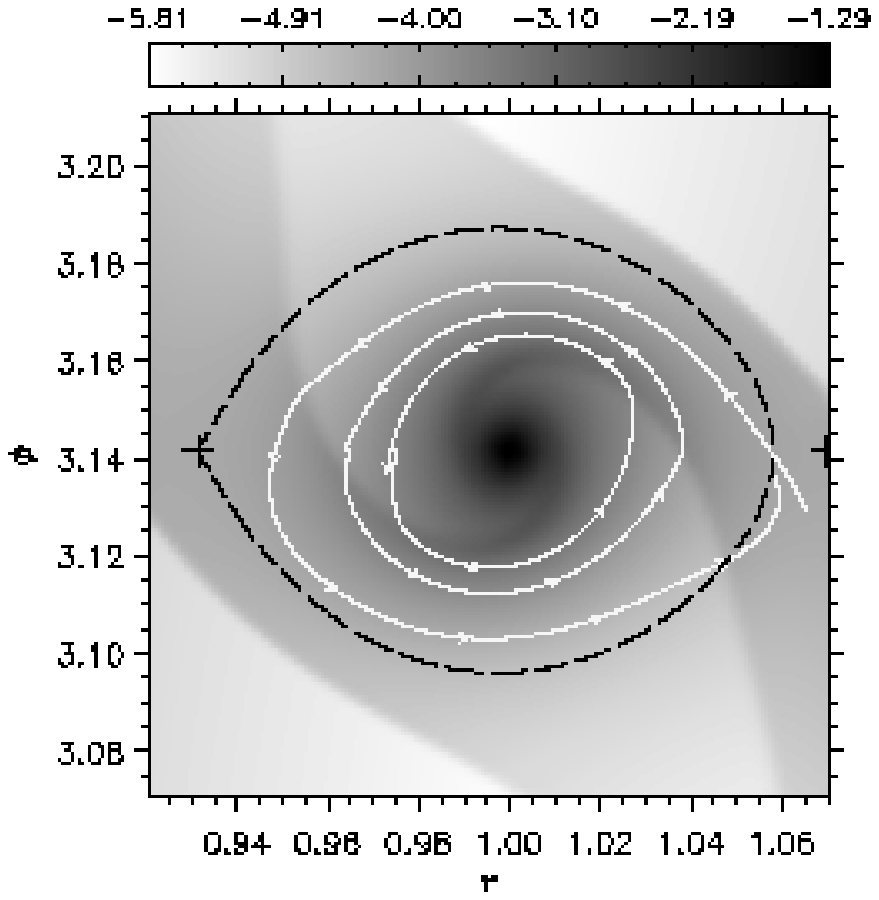}%
\includegraphics{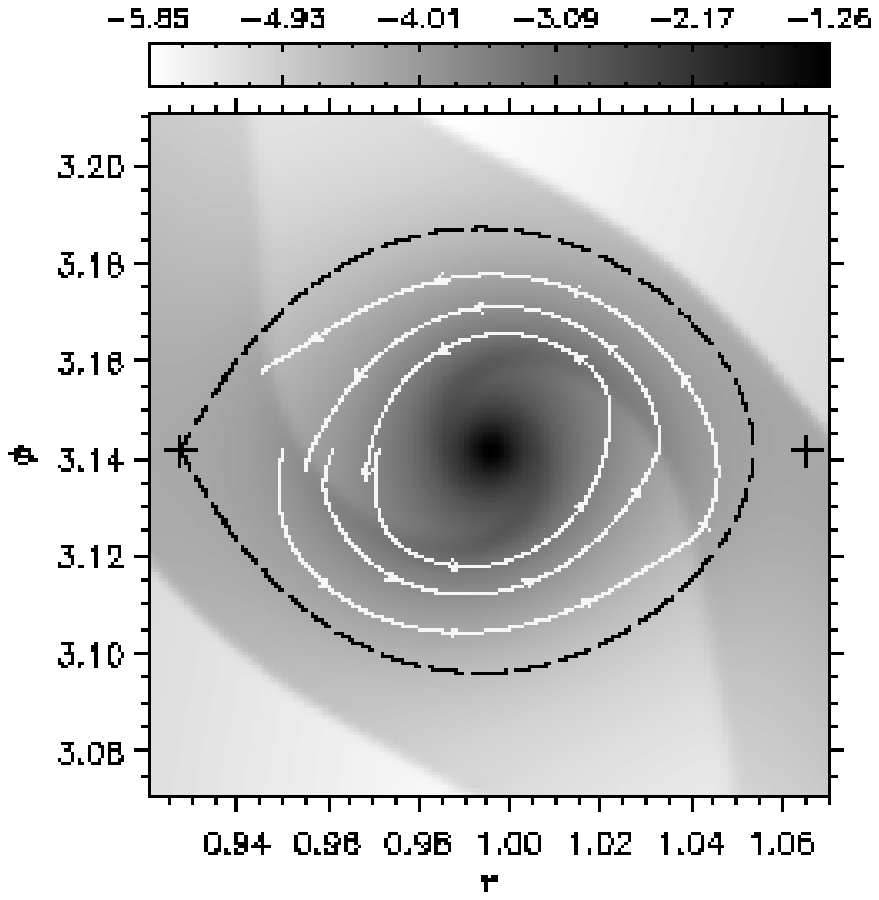}}
\caption{\small%
         Surface density and streamlines around Jupiter-mass, non-accreting
         planets orbiting in a low-mass disc ($\Md=0.01\,\MSun$ within
         $26\,\AU$). 
         The results were obtained from computations executed with the
         grid system 2D4G (linear resolution $\Rhill/\Delta r=38$)
         and softening  $\varepsilon=0.4\,\Rhill$ (\textit{top}) 
         and $0.1\,\Rhill$ (\textit{bottom}). 
         The panels illustrate the situation at the release time $t=300$ 
         orbital periods (\textit{left}) and $50$ orbits later 
         (\textit{right}), 
         while the planet is migrating. The grey-scale is logarithmic and,
         at $5.2\,\AU$, $\Sigma=10^{-3}$ corresponds to $329\,\sdunits$.
         The two streamlines closest to the planet start from distances
         of $\approx 0.36$ and $\approx 0.5\,\Rhill$, respectively.
        }
\label{fig:jup_denstream}
\end{figure*}

Two sets of lines are displayed in each panel of \refFgt{fig:jup_acomp}. 
These are intended to address the lingering question of the importance of 
torques exerted by matter residing deep inside the planet's Hill sphere
\citep{gennaro2003b,bate2003}.
Thus, two configurations were simulated, differing only in whether or not
torques within a radius $\beta \Rhill$ from the planet are included in 
the calculation of the gravitational force in \refeqs{eq:A2}{eq:A1}. 
In one configuration, all torques are taken into account (i.e., $\beta=0$). 
In the second configuration, the simulations were repeated neglecting the 
contribution of material lying inside the inner half of the
Hill sphere (i.e., $\beta=0.5$).  
The choice $\beta=0.5$ was made to avoid the region where the density
gradient is largest (see \refFgp{fig:jup_dengra}) and which mostly contains
material orbiting the planet before it is released 
(see \refFgp{fig:jup_denstream}).

The streamlines in \refFgt{fig:jup_denstream} are constructed by
integrating the velocity field $(u_{r}, u_{\phi})$ at an instant 
in time. Strictly
speaking, this procedure is in error for a migrating planet (right
panels), as it moves during the interval of integration.
But provided the planet moves only a small fraction
of its Hill sphere over the integration time, the streamlines
obtained are reasonably accurate. This condition is satisfied
for the streamlines plotted in this Figure.
It is of course incorrect to ignore torques from within
the Hill sphere since material can move into or out of it (see 
\refFgp{fig:jup_dengra}) and the angular momentum associated with 
this mass flux is lost instead of being transferred to the planet's orbit.
Therefore, migration rates obtained from
configurations with $\beta>0$ are not fully consistent from a physical 
standpoint,
unless one can assure that the neglected material is \textit{constantly}
and not \textit{temporarily} orbiting the planet.

In all cases we considered, the $\beta=0$ calculations resulted in slower 
migration and, thus, these results appear as the upper curves in each panel 
of \refFgt{fig:jup_acomp}. We note that even when torques coming from material 
within the Hill sphere are included, numerical convergence is still achieved. 
This last point turns out to be crucial when high-mass discs are considered 
(see \refSecp{sec:hm_models}).

\subsection[]{Three-dimensional simulations}
\label{sec:lm_3d}
\begin{figure*}
\centering%
\resizebox{\linewidth}{!}{%
\includegraphics{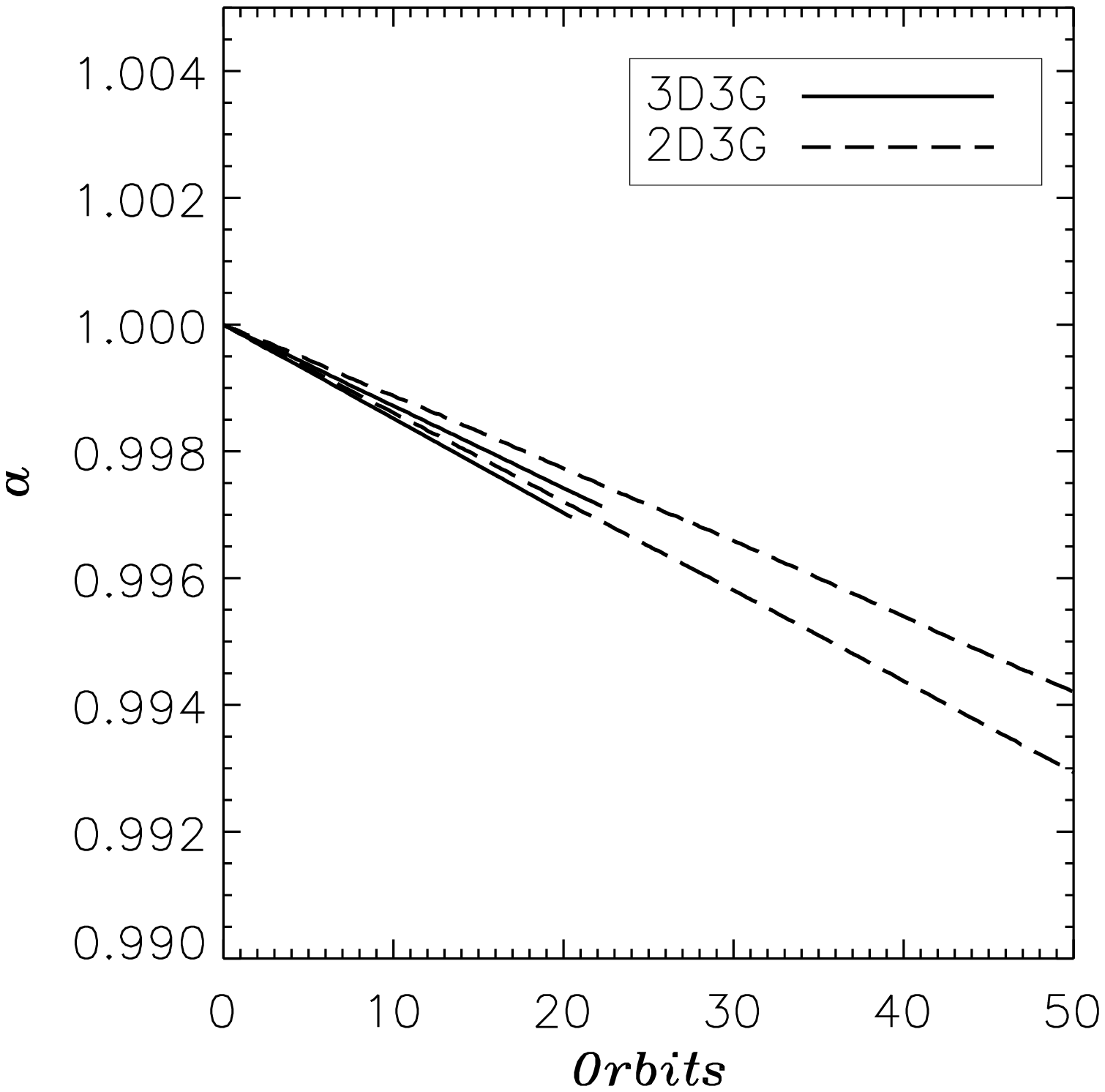}%
\includegraphics{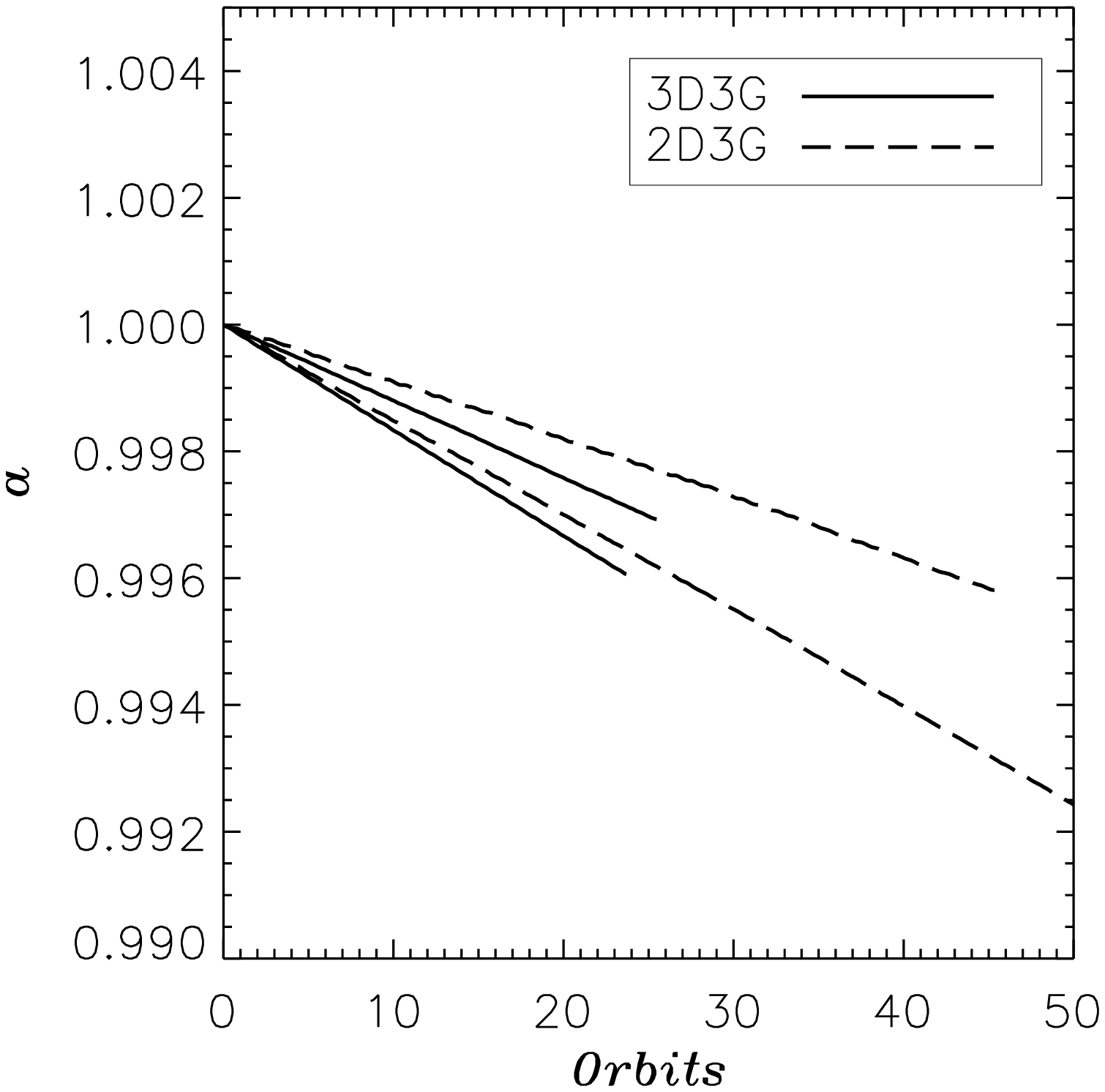}%
\includegraphics{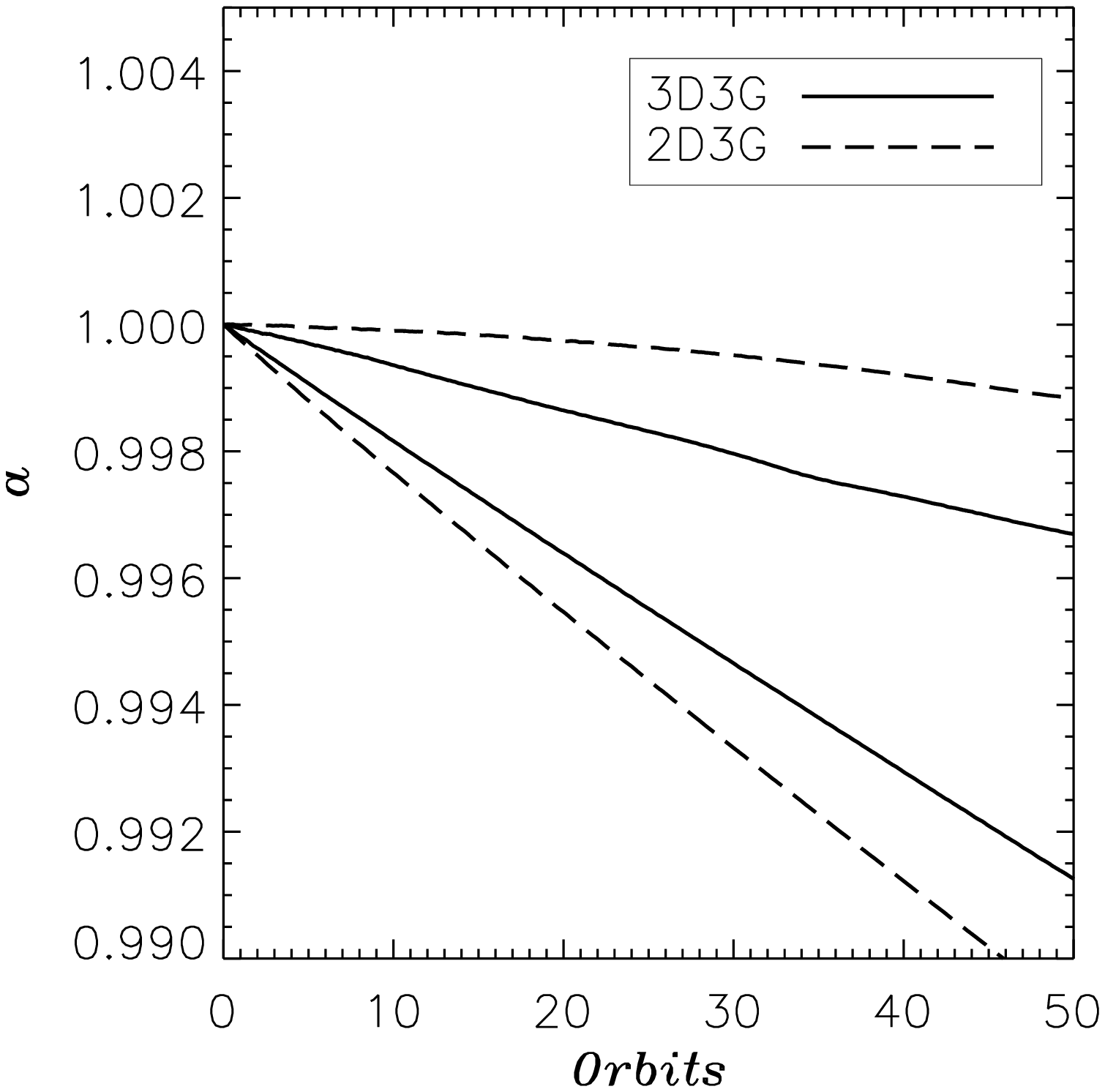}}
\caption{\small%
         Three-dimensional simulations of Jupiter-mass 
         models (grid system 3D3G) orbiting in a low-mass disc,
         compared to analogous two-dimensional models (grid system 2D3G).
         \textit{Left}. Non-accreting planet with 
         gravitational potential softening $\varepsilon=0.4\,\Rhill$.
         \textit{Centre}. Non-accreting planet with 
         $\varepsilon=0.2\,\Rhill$.
         \textit{Right}. Accreting planet with $\varepsilon=0.1\,\Rhill$.
         The release time is equal to $100$ orbits for the non-accreting 
         models and $300$ orbits for the accreting models.
         In each panel, the slower migration occurs with the configuration
         executed with $\beta=0$, whereas the faster migration occurs with
         the configuration run with $\beta=0.5$.}
\label{fig:jup_acomp3d}
\end{figure*}
The vertical stratification of the flow variables in (vertically isothermal) 
discs does not play an important role in determining the strength of Lindblad 
torques acting on Jupiter-mass planets, provided that $\Rhill\ga H$ 
\citep{kley2001}.
However, since the flow structure around the Hill sphere of the planet is 
fully three-dimensional \citep{gennaro2003b,bate2003}, the amount of angular 
momentum delivered by material in the vicinity of the planet may be affected 
by the vertical motion of the fluid. We attempted to investigate this issue 
by means of 3D calculations (grid system 3D3G), whose results are shown in 
\refFgt{fig:jup_acomp3d}. As for 2D computations, the effects of torques 
exerted by material within the Hill sphere were measured by running models 
with $\beta=0$ and $\beta=0.5$. In each panel of \refFgt{fig:jup_acomp3d},
the evolution of the semi-major axis (solid lines) is compared to that 
obtained from 2D models (dashed lines) having an analogous grid system (2D3G).
We were unable to test for convergence of the 3D calculations due to 
computational limitations (simulations with a factor $2$ increase in linear 
resolution would have required around $5000$ CPU hours each). 
But since the 2D 
calculations were converged, we speculate that at the same resolution 3D 
calculations are also converged because the density structure around and 
inside the Hill sphere is smoother in three dimensions.

\begin{figure}
\centering%
\resizebox{1.0\linewidth}{!}{%
\includegraphics{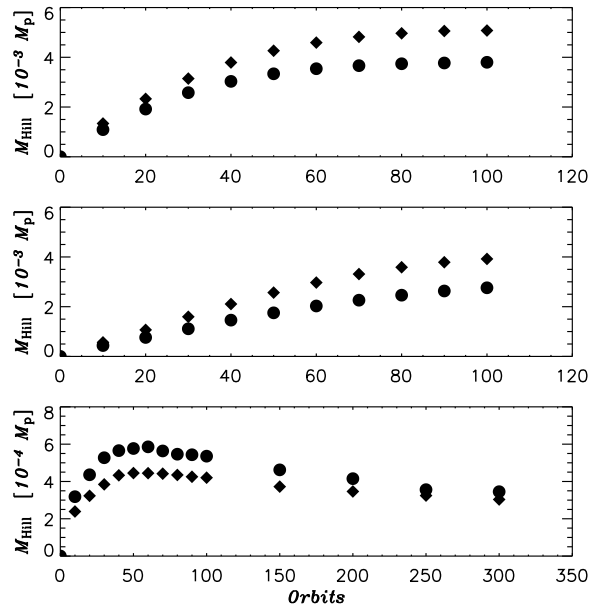}}
\caption{\small%
         Mass enclosed within the Hill sphere of a $\Mp=1\,\MJup$ planet, 
         as measured in 2D (circles) and 3D (diamonds) calculations.
         \textit{Top}. Non-accreting model with  $\varepsilon=0.4\,\Rhill$.
         \textit{Centre}. Non-accreting model with  $\varepsilon=0.2\,\Rhill$.
         \textit{Bottom}. Accreting model with  $\varepsilon=0.1\,\Rhill$.
         Whether or not a gap is imposed on the initial density structure, 
         the amount of material within the Hill sphere tends toward the same 
         value.
         }
\label{fig:jup_rhmass}
\end{figure}

\refFgt{fig:jup_acomp3d} shows that the two- and three-dimensional results 
are similar in the non-accreting cases.  The Hill spheres of non-accreting 
planets contain more material in 3D than they do in 2D (see 
\refFgp{fig:jup_rhmass}). Therefore, it is reasonable to expect that
the migration is slightly faster in three dimensions.

The situation appears more complex in the accreting case, for which migration
is slower in 3D than in 2D if $\beta=0.5$, but it is faster if $\beta=0$ 
(\refFgp{fig:jup_acomp3d}, right panel). 
Since the mass inside the Hill sphere is nearly the same in the two geometries 
(\refFgp{fig:jup_rhmass}, bottom panel), we ascribed this discrepancy to the 
strong spiral waves that occur in the two-dimensional accreting calculations 
\citep{lubow1999,gennaro2002} which are much weaker in three-dimensions due to 
the possibility of vertical motions. Strong spiral waves do not develop when 
$\varepsilon$ is a fair fraction of $\Rhill$.  The non-accreting models with 
$\varepsilon=0.4$ and $0.2\,\Rhill$ present a nearly featureless density 
structure close to the planet in both geometries, hence the similarity of the 
migration rates. 

Finally, we note that non-accreting 3D migration rates with $\varepsilon=0.4$ 
or $0.2\,\Rhill$ are very similar. This suggests that no large variations 
should be expected if smaller values of $\varepsilon$ were employed, provided 
that a sufficiently refined mesh is utilised 
($\Delta R=R\,\Delta\phi\ll\varepsilon$).
The same conclusion seems to be valid in two dimensions, as discussed more 
quantitatively in the next section.

\subsection[]{Migration rates: a quantitative analysis}
\label{sec:quant_analysis}
So long as the semi-major axis does not change significantly (i.e., it 
remains of the same order of magnitude), the migration of a Jupiter-mass 
planet roughly follows an exponential decay
\citep{rnelson2000}. We assume that even when the action of torques arising 
from corotation regions and from material orbiting the planet are included,
the evolution of $a$ can also be described by an exponential decay law
\begin{equation}
  a(t)=a_0\,e^{-(t-t_{\rmn{rls}})/\taum},
\label{eq:alaw}
\end{equation}
for times $t\ge t_{\rmn{rls}}$. 
We take the migration time-scale $\taum=a/|\dot{a}|$ to be a constant over 
the simulated time-interval of the actual planet's migration (between $40$ 
and $100$ orbits). 
This simple parameterisation of $a=a(t)$ is very useful because $\taum$ can 
be directly connected to the acting torques. In fact, if the orbit 
eccentricity is negligible then the conservation of the orbital angular 
momentum leads to the relation 
\begin{equation}
 \dot{a}=\frac{2\,\gvec{\mathcal{T}}\bmath{\cdot}\gOmega}{\Mp\,a\,\Omega^2}, 
 \label{eq:static_adot}
\end{equation}
in which the vector $\gvec{\mathcal{T}}$ denotes the total external torque.
This expression is commonly used to evaluate $\dot{a}$ from the vertical 
component of $\gvec{\mathcal{T}}$ (which we simply indicate as  $\mathcal{T}$),
when a planet moves on a static orbit (i.e., it is not allowed to migrate).

We obtained estimates of $\taum$ for all of the 2D models (grid system 2D4G) 
by performing a linear least-mean-squared fit of the relation 
$\ln{(a/a_0)}=-(t-t_{\rmn{rls}})/\taum$. The results are listed in the second
and third columns of \refTab{tbl:static_moving}, labelled as ``moving'' 
migration time-scale.
The relative error on each estimate is at most $10^{-3}$ and only for this
reason $\taum$ is given with three significant digits.
Discrepancies between estimates computed from 2D and 3D non-accreting models 
are below $\sim10$ per cent. Since the accreting models present a more 
significant discrepancy, $\taum$ is also reported for the simulations in 
three dimensions.

Including the effect of matter orbiting the planet tends to slow down its 
inward drifting motion, regardless of the employed disc geometry, as clearly 
indicated in \refFgt{fig:jup_acomp3d}. The comparison between the $\beta=0$ and 
$\beta=0.5$ migration time-scales shows that the torque from this material can 
be comparable to that from corotation and Lindblad resonances. The total 
(positive) torque produced inside the inner half of the Hill sphere is 
\begin{equation}
 \mathcal{T}_{\rmn{HS}}=\mathcal{T}-\mathcal{T}_{\rmn{LC}}
 \propto\frac{1}{\taum[\beta=0]}-\frac{1}{\taum[\beta=0.5]},
 \label{eq:torque_HS}
\end{equation}
whereas the magnitude of the total (negative) torque exerted from the rest 
of the disc (i.e., Lindblad and corotation torques),
$\left|\mathcal{T}_{\rmn{LC}}\right|$, is proportional to
$1/\taum[\beta=0.5]$. Hence, the ratio between the two contributions is
\begin{equation}
 \frac{ \mathcal{T}_{\rmn{HS}}}{\left|\mathcal{T}_{\rmn{LC}}\right|}%
              =1-\frac{\taum[\beta=0.5]}{\taum[\beta=0]}.
 \label{eq:torque_ratio}
\end{equation}
Entries in the second and third columns of \refTab{tbl:static_moving} indicate 
that in the model with softening $\varepsilon=0.4\,\Rhill$ the material close 
to the planet accounts for a relatively small contribution ($22$ per cent). 
However, shorter smoothing lengths dramatically increase the torque ratio, 
which becomes greater than $60$ per cent in the non-accreting models with 
$\varepsilon=0.2\,\Rhill$ and $0.1\,\Rhill$. A similar ratio between torques 
is obtained in the 3D accreting model.

These migration time-scales can be compared with the Type~I (no gap, resonant)
time-scale of about $4 \times 10^2$
orbits in 2D and $6 \times 10^2$ orbits in 3D \citep{tanaka2002}
and the Type~II (viscous) time-scale $2\,a^2/(3\,\nu) \simeq 10^4$ orbits.

Note that the 2D migration rates tend to converge as $\varepsilon$ is 
decreased.  In particular, the migration rates for $\varepsilon=0.2\,\Rhill$ 
and $0.1\,\Rhill$ differ by less than $10$ per cent. 
\subsection[]{Comparison of migration rates of static and migrating
 planets: the Jupiter-mass case}
\label{sec:mjup_lmd}
\begin{table}
 \begin{center}
 \caption{Comparison of static and moving migration time-scales for a 
          Jupiter-mass planet in a low-mass disc.}
 \label{tbl:static_moving}
 \begin{tabular}{@{}ccccc@{}}
  \hline
 &\multicolumn{2}{c}{Moving}& 
  \multicolumn{2}{c}{Static}\\
  \cline{2-3}\cline{4-5}
  \raisebox{1.5ex}[-1.5ex]{$\varepsilon$} & $\beta=0$   & $\beta=0.5$   
                                          & $\beta=0$   & $\beta=0.5$ \\
  \hline
    $0.4\,\Rhill^{\phantom{\dag}}$   
                    & $9.94\times10^{3}$  & $7.75\times10^{3}$
                    & $1.0\times10^{4}$   & $7.8\times10^{3}$         \\
    $0.2\,\Rhill^{\phantom{\dag}}$   
                    & $1.76\times10^{4}$  & $6.34\times10^{3}$
                    & $1.8\times10^{4}$   & $6.2\times10^{3}$         \\
    $0.1\,\Rhill^{\phantom{\dag}}$   
                    & $1.51\times10^{4}$  & $5.71\times10^{3}$
                    & $1.8\times10^{4}$   & $5.7\times10^{3}$         \\
    $0.1\,\Rhill^{\dag}$       
                    & $5.97\times10^{4}$  & $4.86\times10^{3}$
                    & $4.8\times10^{4}$   & $4.2\times10^{3}$         \\
    $0.1\,\Rhill^{\ddag}$       
                    & $1.54\times10^{4}$  & $5.77\times10^{3}$
                    & $1.5\times10^{4}$   & $5.4\times10^{3}$         \\
  \hline
  \multicolumn{5}{l}{$^{\dag}$ 2D accreting model. $^{\ddag}$ 3D accreting model.} \\
 \end{tabular}
 \end{center}
\small{%
The migration time-scales labelled as ``moving'' refer to the time-scale, 
$\taum$, in \refeqt{eq:alaw} and were computed as explained in 
\refSect{sec:quant_analysis}.
They are expressed in units of initial orbital periods, i.e., $11.9$ years 
if $a_0=5.2\,\AU$. One-standard deviation uncertainties for these estimates 
range from $1$ to $10$ orbits. See \refSect{sec:lm_cs} for an explanation 
of configurations $\beta=0$ and $\beta=0.5$.
Migration time-scales labelled as ``static'' were determined from  
\refeqt{eq:static_adot} by employing torques averaged over the last ten 
orbits before the release time.
Computations were executed with the grid system 2D4G.
      }
\end{table}
We examined whether the torque exerted on the planet by the disc material is 
influenced by the radial motion of the planet. As discussed in 
\refSect{sec:introduction}, the motion of the planet might be able to affect
the coorbital torques and therefore the migration rate. In order to test
this hypothesis, we computed the total torque acting on the planet during the 
last ten orbital periods before it was released. This was done for both 
$\beta=0$ and $\beta=0.5$ configurations. Since no angular momentum is actually
extracted from or added to the planetary orbit, which thus remains static, 
we shall refer to such torques as \textit{static} torques.
The migration time-scales listed in the two right-most columns of  
\refTab{tbl:static_moving} were obtained from the average static torques by using 
\refeqt{eq:static_adot}. In \refTab{tbl:static_moving} we compared these ``static'' 
migration time-scales, $\taum^{\rmn{S}}$, with the migration time-scales, $\taum$, 
measured from the moving planet calculations. 
In all cases, there is close agreement between the static and moving migration 
time-scales. These results show that under these circumstances of disc and 
planetary masses, there is no strong dependence of the torques on whether 
planets are on fixed orbits or allowed to migrate.

\section[]{Results of high-mass disc models}
\label{sec:hm_models}
The Type~II migration rate depends only on the viscous time-scale of the disc 
near the location of the planet and is independent of the disc density, 
provided that the gap is devoid of material. 
Yet gaps are generally not completely cleared and the Type~II time-scale 
prediction does not take into consideration the angular
momentum exchanged between the planet and the ``gap'' material. Some of this 
material travels on horse-shoe orbits, while other material circulates within 
the planet's Hill sphere. The angular momentum delivered in either case may 
play a major role in planetary migration \citep[see, e.g.,][]{masset2001} and 
it is proportional to the local mass density. In fact, MP03 recently claimed 
that there exists a critical mass (when the material around the planet is more 
massive than the planet), beyond which a runaway migration process sets in.

\begin{figure}
\centering%
\resizebox{1.0\linewidth}{!}{%
\includegraphics{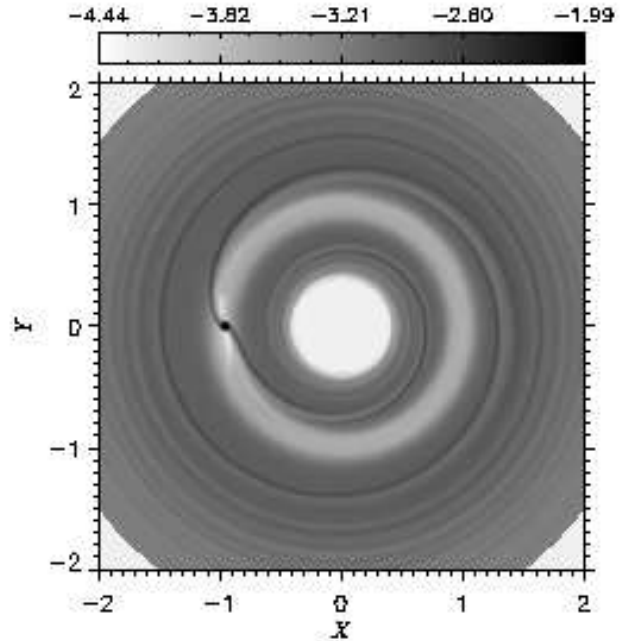}}
\caption{\small%
         Global surface density around a $\Mp=0.3\,\MJup$
         (non-accreting) planet orbiting in a $0.024\,\MSun$ disk.
         The density is displayed at $t=550$ orbits,
         when the planet has migrated for about $70$ orbits. 
         The grey-scale is logarithmic and $10^{-3}$ corresponds to
         $329\,\sdunits$, at $5.2\,\AU$. The average gap density
         is $2.5\times10^{-4}$ or $82\,\sdunits$. 
        }
\label{fig:map_global}
\end{figure}
We ran simulations of Saturn-like bodies ($\Mp=0.3\,\MJup$) embedded in a disc 
as massive as $24\,\MJup$ inside $13\,\AU$. The annular region within 
$2\,\Rhill$ from the planet is initially $7.5$ as massive as the planet. 
Nonetheless, the aspect ratio is small enough ($H/r=0.03$) so the thermal 
condition for gap formation, $\Mp/\MStar > 3\,(H/r)^3$ 
\citep[e.g.,][]{lin1993}, is fulfilled and therefore the migration might be 
within the Type~II regime, although the gap is not completely cleared
as can be seen in \refFgt{fig:map_global}. In 
these cases, with massive discs and small aspect ratios, very large density 
gradients develop inside the Hill sphere. Therefore, 
it is especially important to investigate the dependence of the results on 
numerical resolution. We achieved convergence for the flow outside
of the Hill sphere by using numerical resolutions of order $13$ grid zones
per Hill radius. However, in order to accurately determine the contributions 
to the migration rate from material inside the Hill sphere, resolutions
higher than $52$ grid zones per Hill radius are necessary.
\subsection[]{Convergence tests}
\label{sec:hm_ct}
\begin{figure*}
\centering%
\resizebox{0.90\linewidth}{!}{%
\includegraphics{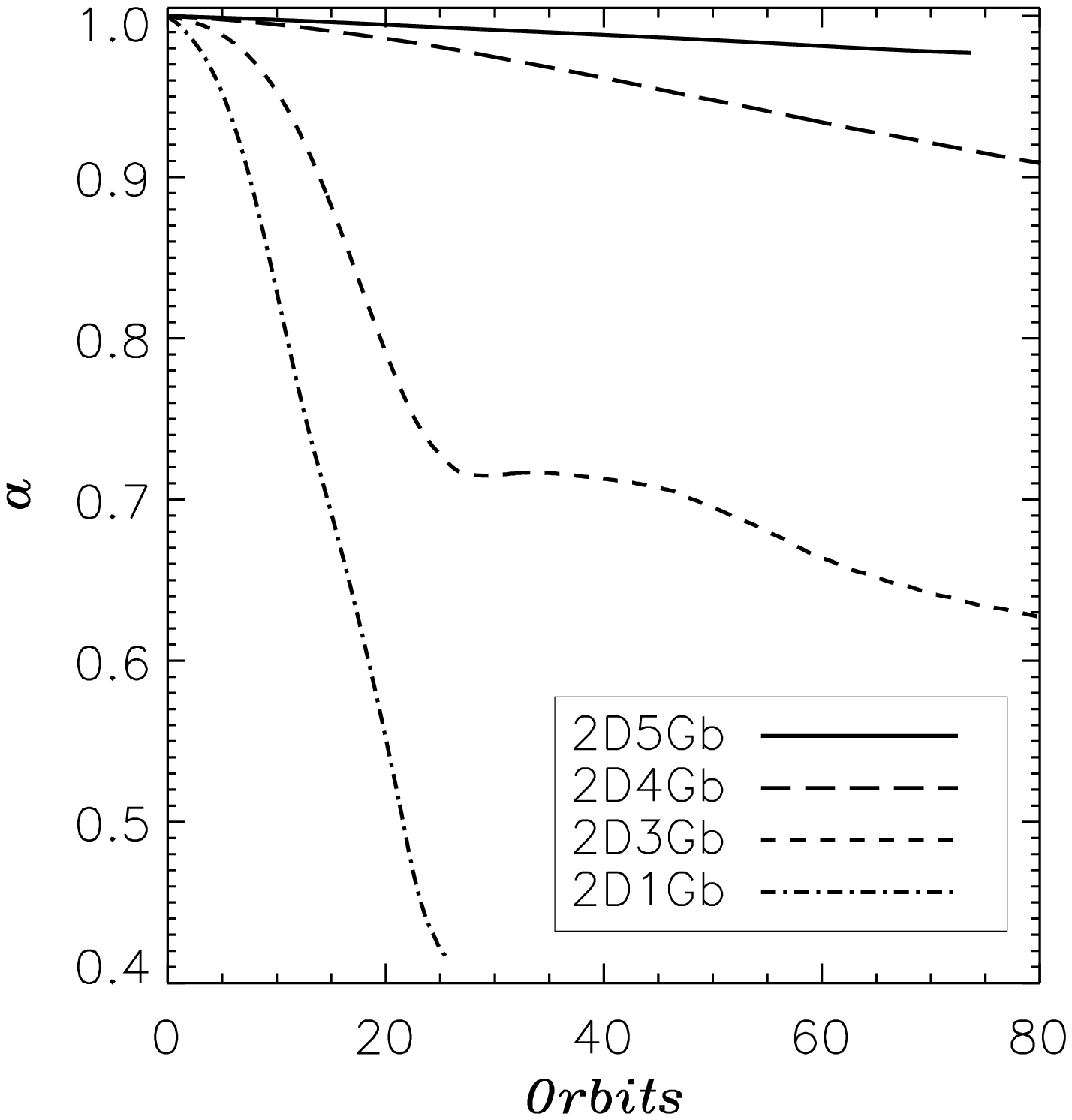}%
\includegraphics{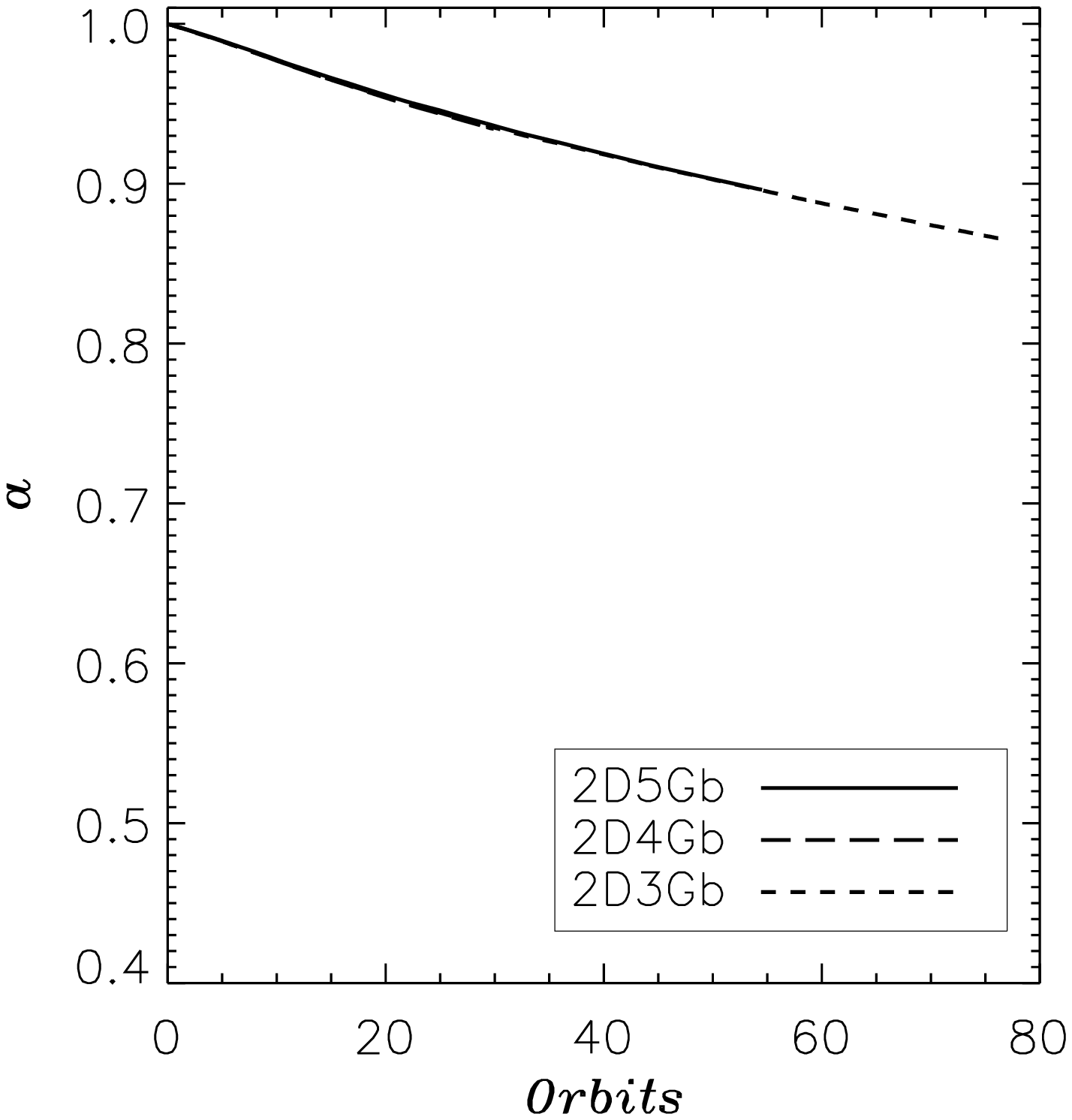}}
\caption{\small%
         Computations of a Saturn-mass planet orbiting in a high-mass disc:
         convergence tests.
         The gravitational potential softening, $\varepsilon$, is $60$
         per cent of the local disc scale-height.
         The release time is equal to $477$ orbits.
         The left panel shows the evolution of $a$ when all torques are
         taken into account (i.e., $\beta=0$). The right panel shows how the
         evolution of the semi-major axis proceeds when the contribution of 
         those
         torques arising from inside the Hill sphere (i.e., $\beta=1.0$) 
         are neglected .
         The dot-dash line (labelled as 2D1Gb) refers to a calculation 
         executed with the same numerical resolution as in MP03.
         }
\label{fig:map_acomp}
\end{figure*}
As mentioned in \refSect{sec:hm_par}, the model setup and the disc parameters
were chosen to match as closely as possible those in MP03. The smoothing length
was $60$ per cent of the local disc thickness, $H$, 
(i.e., $\varepsilon=0.3878\,\Rhill$), the planet was non-accreting, and
$t_{\rmn{rls}}=477$ orbits. We performed a calculation using a 
single-level grid (2D1Gb, see \refTab{tbl:grids2}) aimed at reproducing 
the resolution used by MP03 ($\Delta r/\Rhill\simeq 0.3$ and 
$\Delta r/\varepsilon\simeq 0.8$). 
We then performed a convergence test using different numerical resolutions, as 
provided by the grid systems 2D3Gb, 2D4Gb, and 2D5Gb (see \refTab{tbl:grids2}).
An additional convergence test, involving the grid system 2D6Gb, is discussed
in \refSect{sec:map_conv}.

The left panel of \refFgt{fig:map_acomp} shows the outcomes of the tests for 
the evolution of the semi-major axis concerning the configuration with 
$\beta=0$. The dot-dash line in this panel represents the result from the 
single-grid computation 2D1Gb, which should be compared to the model 
labelled as $\rmn{S}_{8}$ in Figure~2 of MP03. Given the remarkable 
agreement between our and their outcome, we are confident that we
reproduced the same physical and numerical conditions for runaway migration. 
Yet, computations repeated with finer and finer resolutions gave smaller and 
smaller migration rates which, as displayed in \refFgt{fig:map_acomp} (left 
panel), failed to converge. The gain in linear resolution achieved (over the 
single-grid simulation) with the employed grid systems ranges from $4$ (2D3Gb) 
to $16$ (2D5Gb). In the highest resolution models, there are $52$ grid zones 
per Hill radius. Comparing the short-dash and dot-dash curves in left panel of 
\refFgt{fig:map_acomp}, one realises that the average migration speed obtained 
over the first $25$ orbits with the grid system 2D3Gb is only half (in physical units, 
$\langle\dot{a}\rangle\approx -5\times10^{-3}\,\AU\,\rmn{yr}^{-1}$) 
of that in MP03.
Calculations executed with the grid systems 2D4Gb and 2D5Gb give even lower 
migration speeds of $\langle\dot{a}\rangle\simeq -5\times10^{-4}$ and 
$-1.4\times10^{-4}\,\AU\,\rmn{yr}^{-1}$, respectively.
 
While there is a factor of $10$ decrease in disc torques acting on
the planet in going from grid systems 2D3Gb to 2D4Gb, this factor reduces
to $3.6$ when the two most refined grid systems are considered.
Yet, from the behaviour of semi-major axis evolution shown in the left panel 
of \refFgt{fig:map_acomp}, we cannot determine whether it is converging.
To assess this point we employed the grid system 2D6Gb 
(see \refSecp{sec:map_conv})
which indicates that the solid line in \refFgt{fig:map_acomp} is basically a 
converged evolution.

\begin{figure}
\centering%
\resizebox{1.0\linewidth}{!}{%
\includegraphics{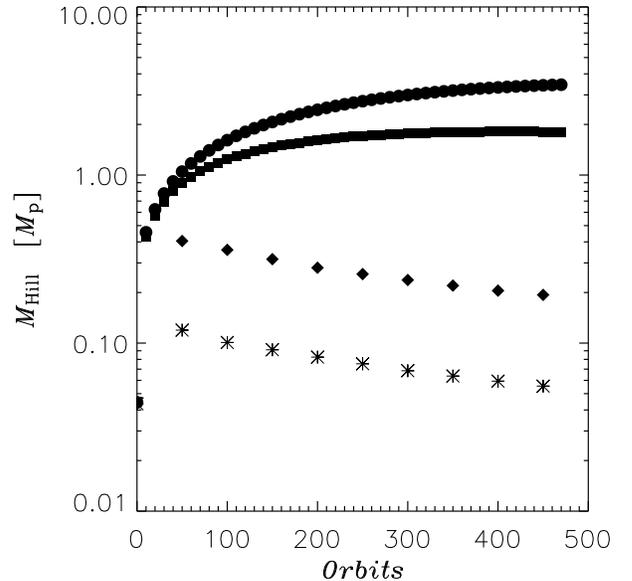}}
\caption{\small%
         Mass enclosed within the Hill sphere of a $\Mp=0.3\,\MJup$ planet, 
         as measured from simulations with increasing resolutions:
         single-level grid 2D1Gb (\textit{asterisks});
         grid system 2D3Gb (\textit{diamonds});
         grid system 2D4Gb (\textit{squares});
         grid system 2D5Gb (\textit{circles}).
         }
\label{fig:map_rhmass}
\end{figure}
\begin{figure}
\centering%
\resizebox{1.0\linewidth}{!}{%
\includegraphics{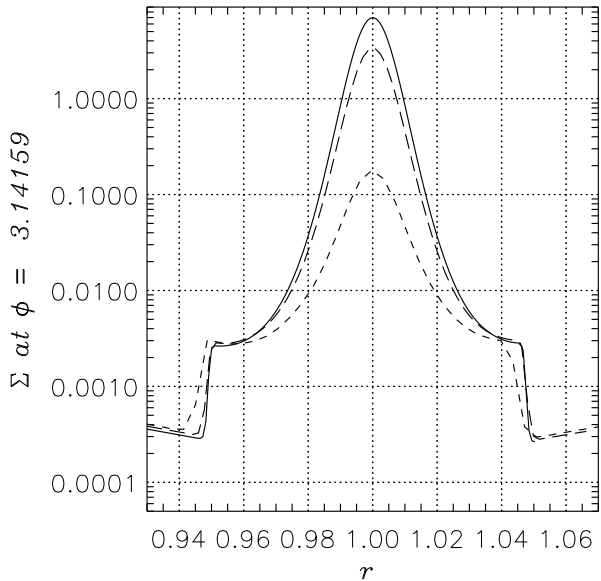}}
\caption{\small%
         Surface density profile at the azimuthal position 
         of a planet with $\Mp/\MStar=3\times10^{-4}$,
         after $450$ orbital periods (the planet starts migrating at 
         $t_{\rmn{rls}}=477$ orbits), 
         orbiting within a high-mass disc.
         Different line types refer to computations performed with 
         different grid systems:
         2D5Gb (\textit{solid line});
         2D4Gb (\textit{long-dash line});
         2D3Gb (\textit{short-dash line}).
         If $a_{0}=5.2\,\AU$ and $\MStar=1\,\MSun$, $\Sigma=10^{-2}$ 
         is equal to $3.29\times10^{3}\,\sdunits$. 
        }
\label{fig:map_dengra}
\end{figure}
The right panel in \refFgt{fig:map_acomp} shows the semi-major axis evolution 
from the same calculations as in the left panel but executed with $\beta=1.0$, 
i.e., excluded torques arising inside the planet's Hill sphere.
As before, this choice of $\beta$ was made to exclude the region  around the 
planet with largest density gradients as well as largest torque densities.
Clearly, numerical convergence was readily achieved with this configuration.
The migration time-scale, obtained from a least-mean-squared fit to
the data (see \refSecp{sec:quant_analysis}), is $\taum=493$ orbits.
Furthermore, outcomes of simulations executed with $\beta=0.75$ attained 
convergence at almost the same rate of migration as with $\beta=1.0$.
Therefore, we conclude that the material close to the planet
must be held responsible for the non-convergence of the $\beta=0$ 
configuration in the left panel of \refFgt{fig:map_acomp}. 
That is, the torque arising from
within the Hill sphere converges very slowly with increasing resolution.
Despite the fact that the amount of material inside the planet's Hill
sphere increases as the grid resolution is raised (see  
\refFgp{fig:map_rhmass}), the resulting migration rates or net torques are 
actually smaller.
\refFgt{fig:map_dengra} shows the surface density near the planet at an 
advanced time, shortly before it is allowed to migrate.  
We note that the mass near the planet seems to be converging at the highest 
resolutions, but convergence is not yet formally achieved.
This Figure also implies that most of the material is piled up very
close to the planet. We measured that $80$ per cent of the mass contained
inside the Hill sphere is concentrated within a distance of $0.2\,\Rhill$
from the planet. 

The mass build up within the Hill sphere appears to suggest that disc
self-gravity may be dynamically important. However, this may not
actually be the case. A simple calculation of a viscous disc that
accretes at the typical rates of $10^{-8}\,\MSun$ per year suggests
that it is likely not self-gravitating (the value of the Toomre parameter
$Q$ is much greater than unity for the parameters in this paper).  
In the simulations presented here, the mass build up is concentrated 
in a region of order the smoothing length (see \refFgp{fig:map_dengra}).  
Within that radius, further inward viscous accretion is artificially 
slow because the gravitational potential of the planet tends to enforce 
rigid rotation (see second term in \refeqp{eq:phi}).  
In addition, the boundary condition of no accretion on the planet prevents 
the accumulated gas from being removed from the simulation. 
As such, much of the gas accumulated  within a smoothing length represents
material that is incorporated by the planet, rather than residing in
the disc.

\begin{figure*}[t!]
\centering%
\begin{minipage}[c]{0.90\linewidth}
\centering%
\resizebox{\linewidth}{!}{%
\includegraphics{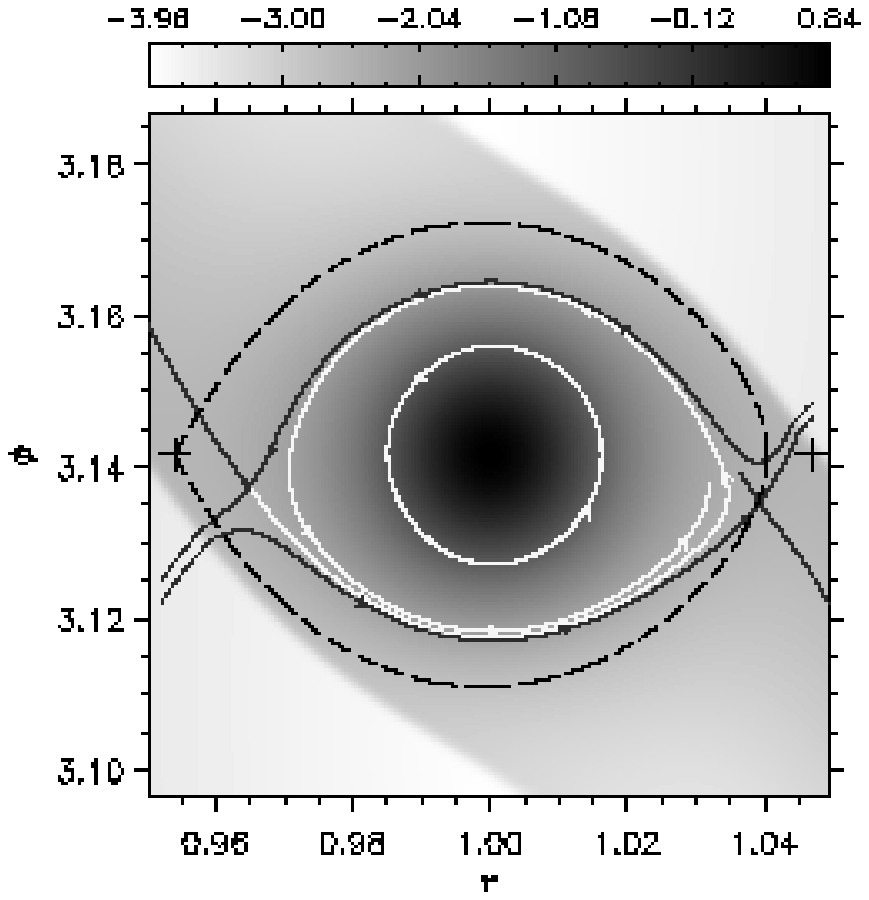}%
\includegraphics{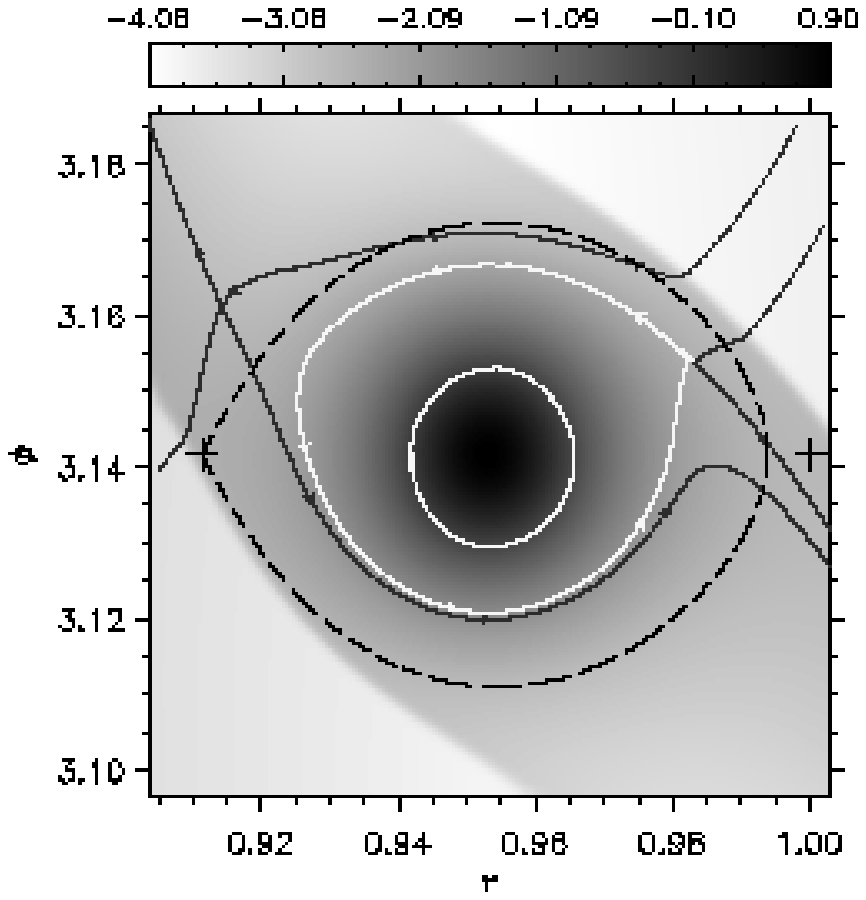}}
\end{minipage}
\hspace*{\fill}%
\begin{minipage}{0.45\linewidth}
\centering%
\resizebox{\linewidth}{!}{%
\includegraphics{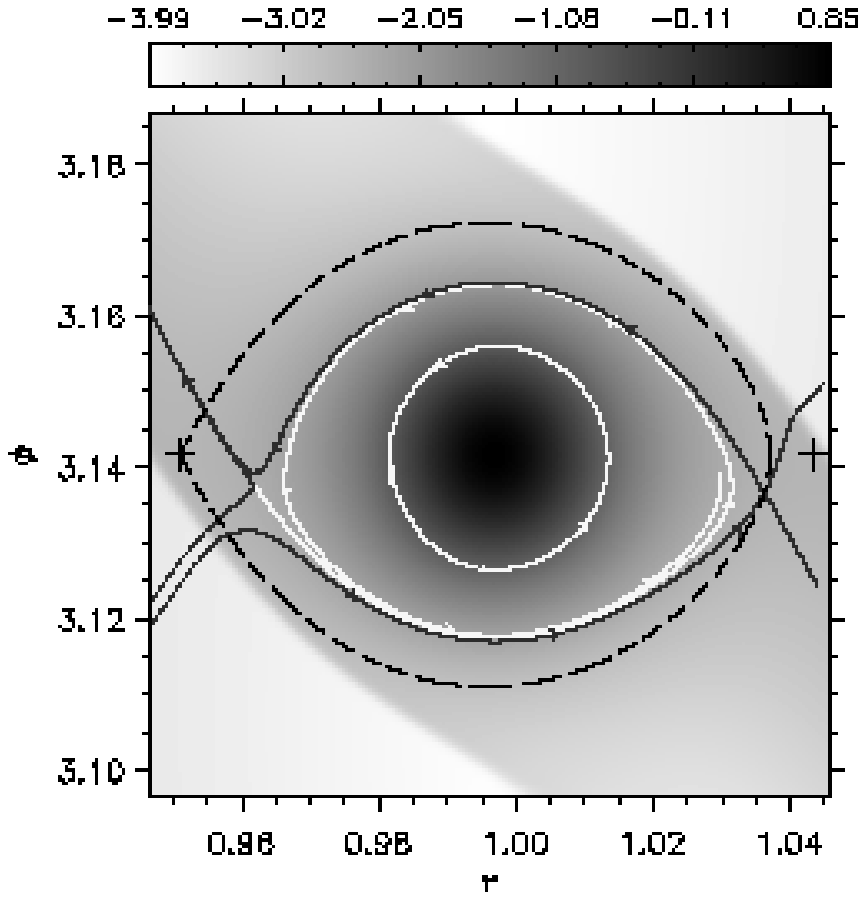}}
\end{minipage}\hfill%
\begin{minipage}{0.40\linewidth}
\centering%
\caption{\footnotesize%
         Surface density and streamlines around a $\Mp=0.3\,\MJup$, 
         non-accreting planet orbiting in a high-mass disc 
         ($\Md=0.024\,\MSun$ within $13\,\AU$).
         The dashed line indicates the Roche lobe and the crosses mark the
         position of the L1 and L2 Lagrange points.
         The top-left panel refers to the time $t=450$ orbital periods (i.e.,
         before it is released) whereas the top-right panel displays
         the situation at $t=497$ orbits with the configuration $\beta=1.0$ 
         (its instantaneous radial speed is 
         $\dot{a}\simeq -1.5\times10^{-4}\,\AU\,\rmn{yr}^{-1}$). 
         The bottom panel refers to the same time but when all
         torques are taken into account (i.e., $\beta=0$).
         The grey-scale is logarithmic and, at $5.2\,\AU$,  
         $10^{-3}$ corresponds to $329\,\sdunits$.
         These results were obtained from computations executed with the
         grid system 2D5Gb (linear resolution
         $\Delta r/\Rhill\simeq 2\times10^{-2}$). 
         As in \refFgt{fig:jup_denstream}, the streamlines in the 
         bottom panel do not account for the planet's motion
         because of the small $\dot{a}$. A more strict procedure
         was instead employed to calculate the streamlines in the 
         top-right panel. In this case we integrated the velocity field 
         $(u_{r}-\dot{a}, u_{\phi})$, where $\dot{a}$ is the 
         instantaneous radial speed of the planet.  
        }
        \label{fig:map_denstream}
\end{minipage}\hspace*{\fill}%
\end{figure*}
Although the configuration with $\beta=1.0$ provides numerically converged
migration rates, one has to be wary of their physical meaning. 
\refFgt{fig:map_denstream} illustrates that before the planet is released 
(top-left panel) material on horse-shoe orbits passes through the Roche lobe
as close to the planet as $\approx 0.5\,\Rhill$. Yet, if the planet starts
rapidly migrating this picture is bound to change. The top-right panel shows
a snapshot after $20$ orbits from the release time, as the planet radially
moves at a speed 
$\dot{a}\simeq -1.5\times10^{-4}\,\AU\,\rmn{yr}^{-1}$ 
(configuration $\beta=1.0$ and grid system 2D5Gb).
The situation appears less symmetric than before the release and the
flow structure within the Hill sphere has been altered by the
rapid planetary motion.
As a reference, we also show in the bottom panel what happens when all
torques are consistently taken into account ($\beta=0$).
We calculated the torques arising from the Hill sphere in the
situation depicted in the top-right panel and we found that
they are three times as large (and more positive) as those exerted, at
the same time, in the configuration $\beta=0$ (bottom panel).
This difference may indicate that the faster motion in the
$\beta=1$ case has artificially
changed the density distribution inside the Hill sphere and thus the 
circulation in the coorbital region.

\begin{figure*}[t!]
\centering%
\resizebox{0.90\linewidth}{!}{%
\includegraphics{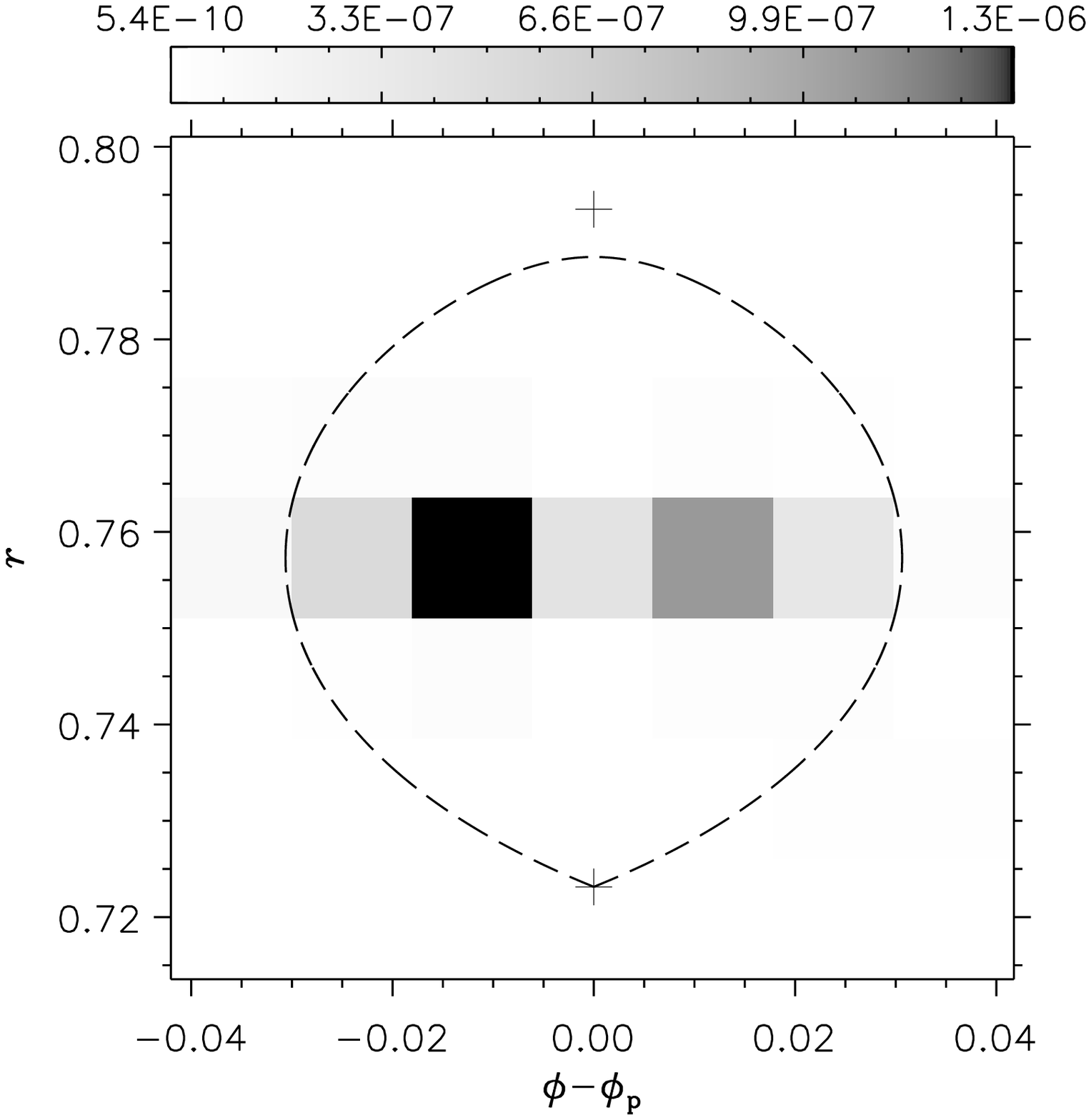}%
\includegraphics{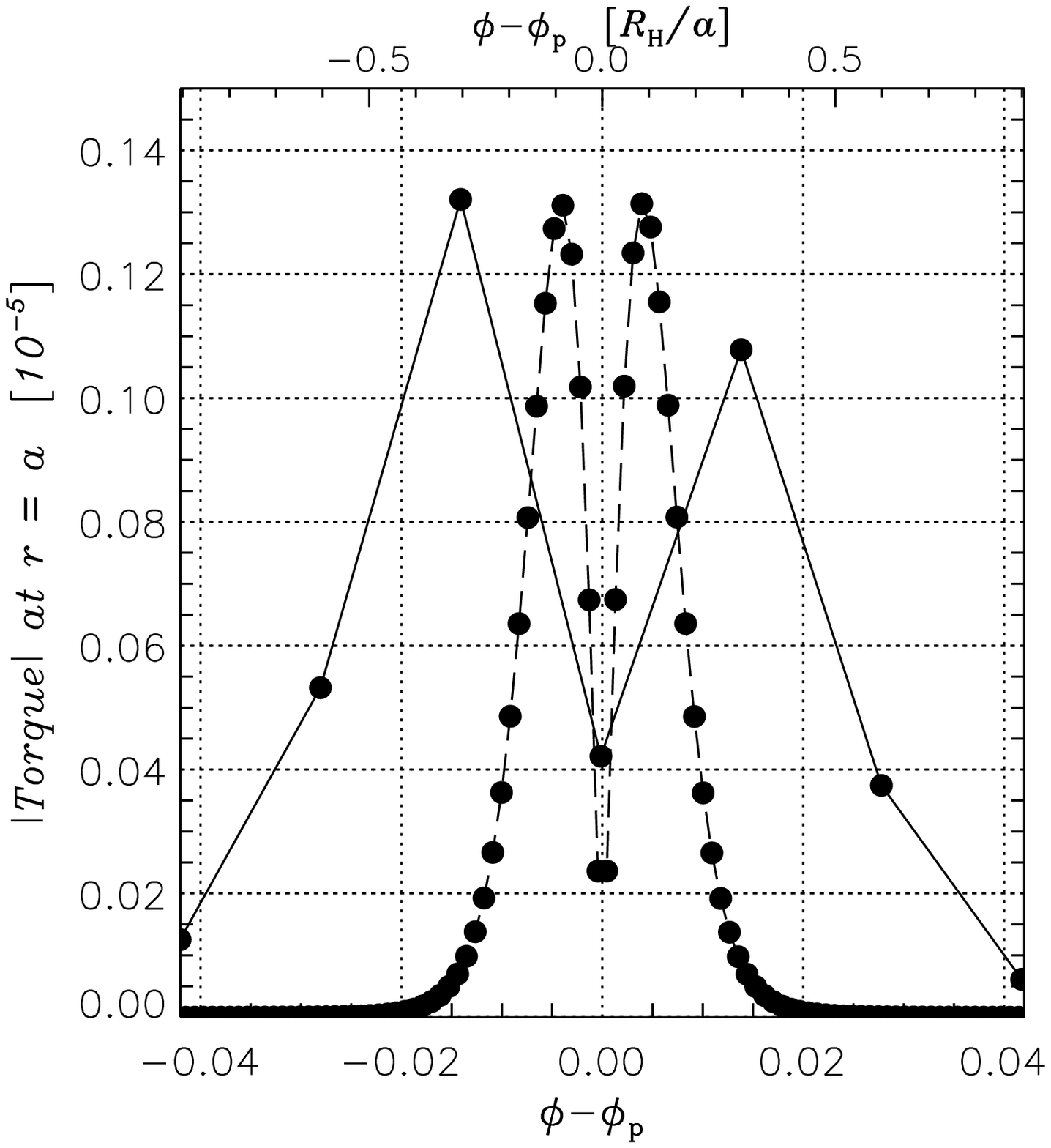}}
\caption{\small%
         \textit{Left}. Absolute value of the torque density (linear scale) 
         close to a planet with $\Mp/\MStar=3\times10^{-4}$ and
         orbiting in a high-mass disc.
         The dashed line indicates the  Roche lobe and the crosses mark the
         L1 and L2 Lagrange points. This calculation was executed with
         the single-level grid 2D1Gb ($\Delta r\simeq 0.3\,\Rhill$).
         Shaded pixels represent the actual size of the grid zones.
         The torque distribution is illustrated at $t=496$ orbital periods
         while the planet is migrating at an average rate
         $\langle\dot{a}\rangle\approx -10^{-2}\,\AU\,\rmn{yr}^{-1}$.
         The torque density is negative when $\phi<\phi_{\rmn{p}}=\pi$ 
         and positive when $\phi>\phi_{\rmn{p}}$.
         It is evident that at such low resolution there is an unbalanced 
         inward torque that is not observed in higher resolution
         calculations (see top panels of \refFgp{fig:map_torden}).
         \textit{Right}. The solid line shows the profile of the absolute 
         value of the torque density (shown in the left panel) through the
         radial position of the planet. The filled circles indicate the
         positions of the data. The dashed line refers to an analogous
         profile from the highest resolution calculation (grid system
         2D5Gb), which was rescaled so that its maximum value was 
         $1.3\times 10^{-6}$.
         }
\label{fig:map_tordenst}
\end{figure*}
\begin{figure*}[t!]
\centering%
\resizebox{0.90\linewidth}{!}{%
\includegraphics{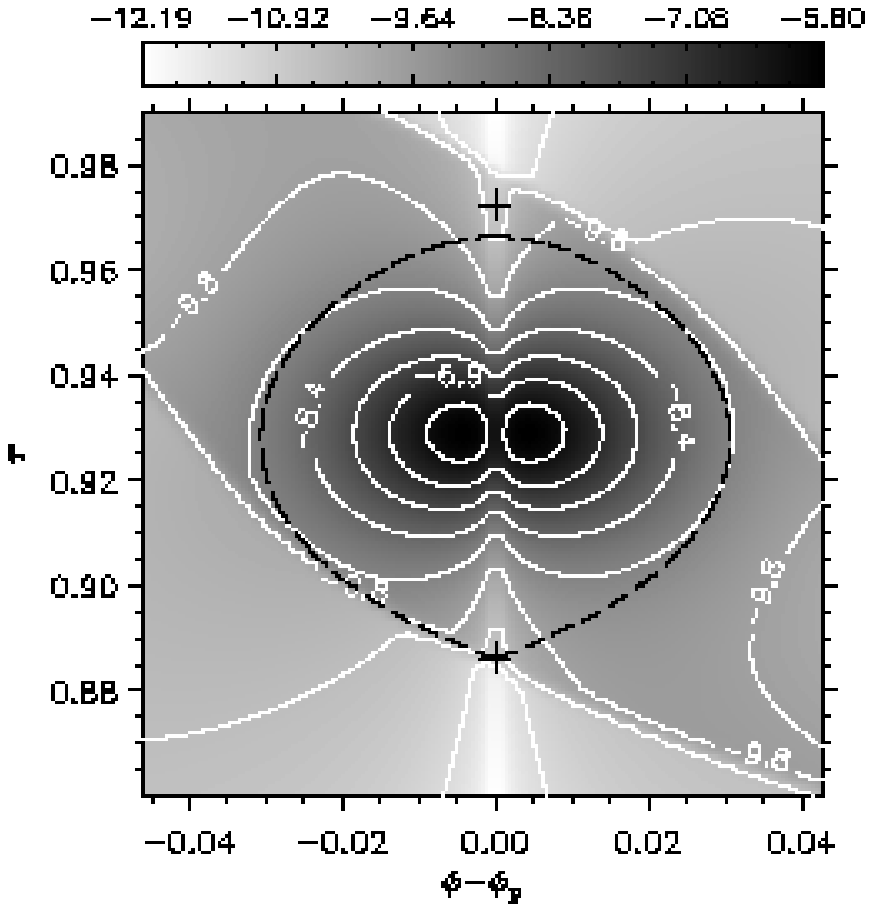}%
\includegraphics{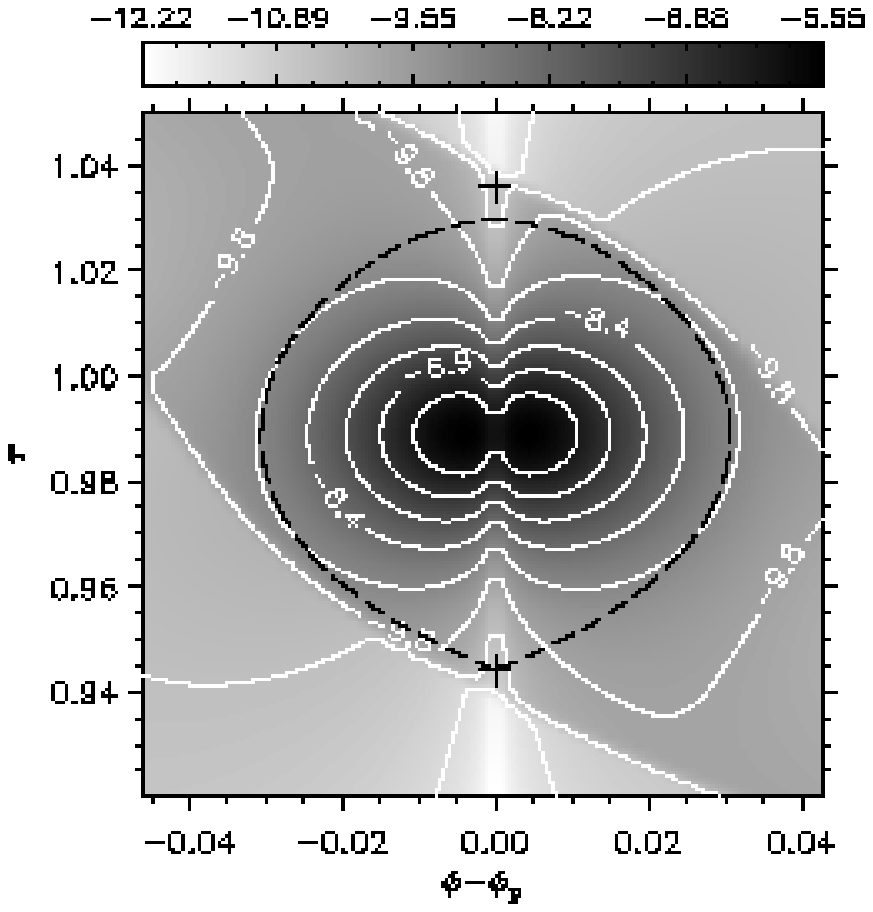}}
\resizebox{0.90\linewidth}{!}{%
\includegraphics{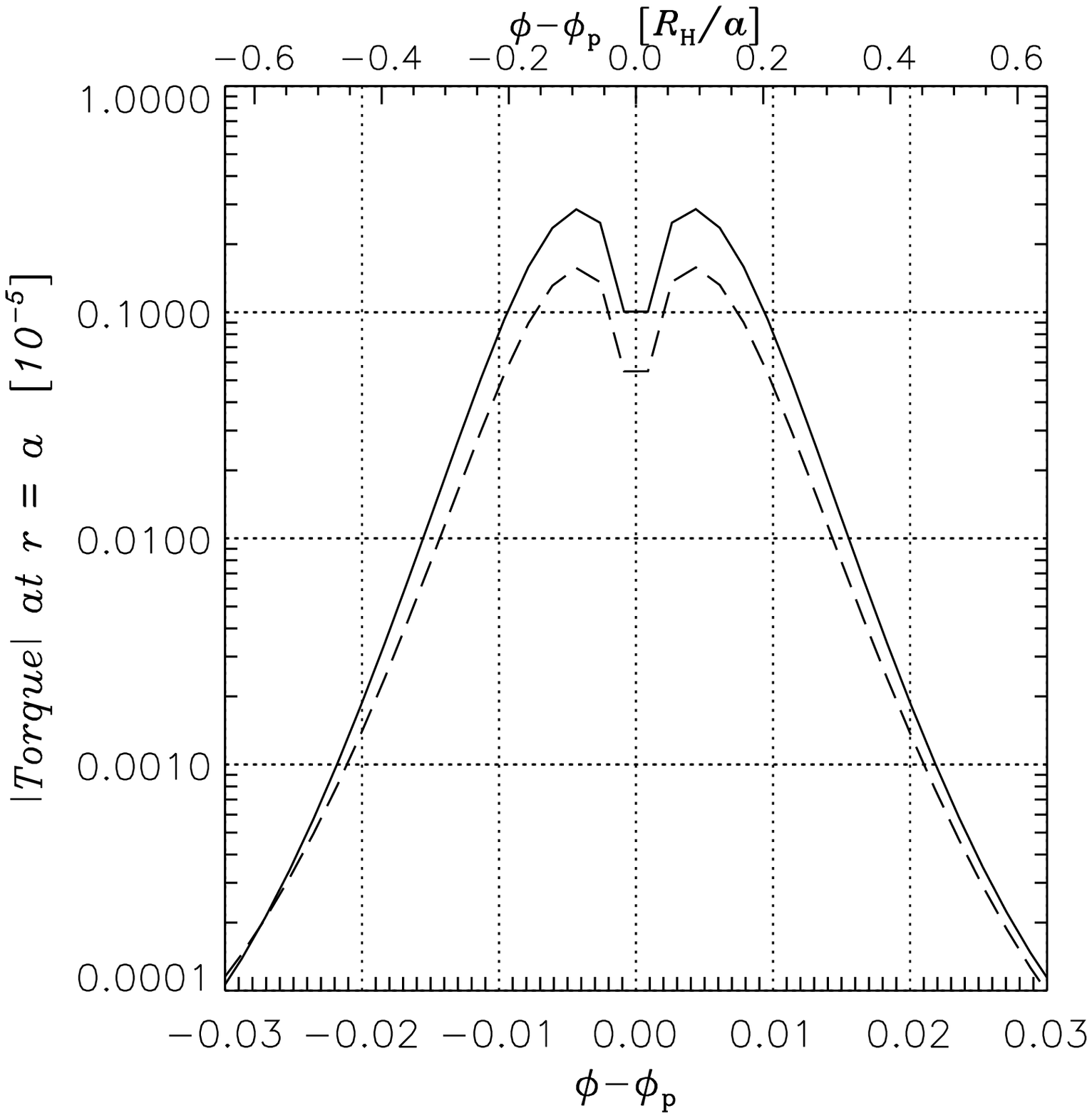}%
\includegraphics{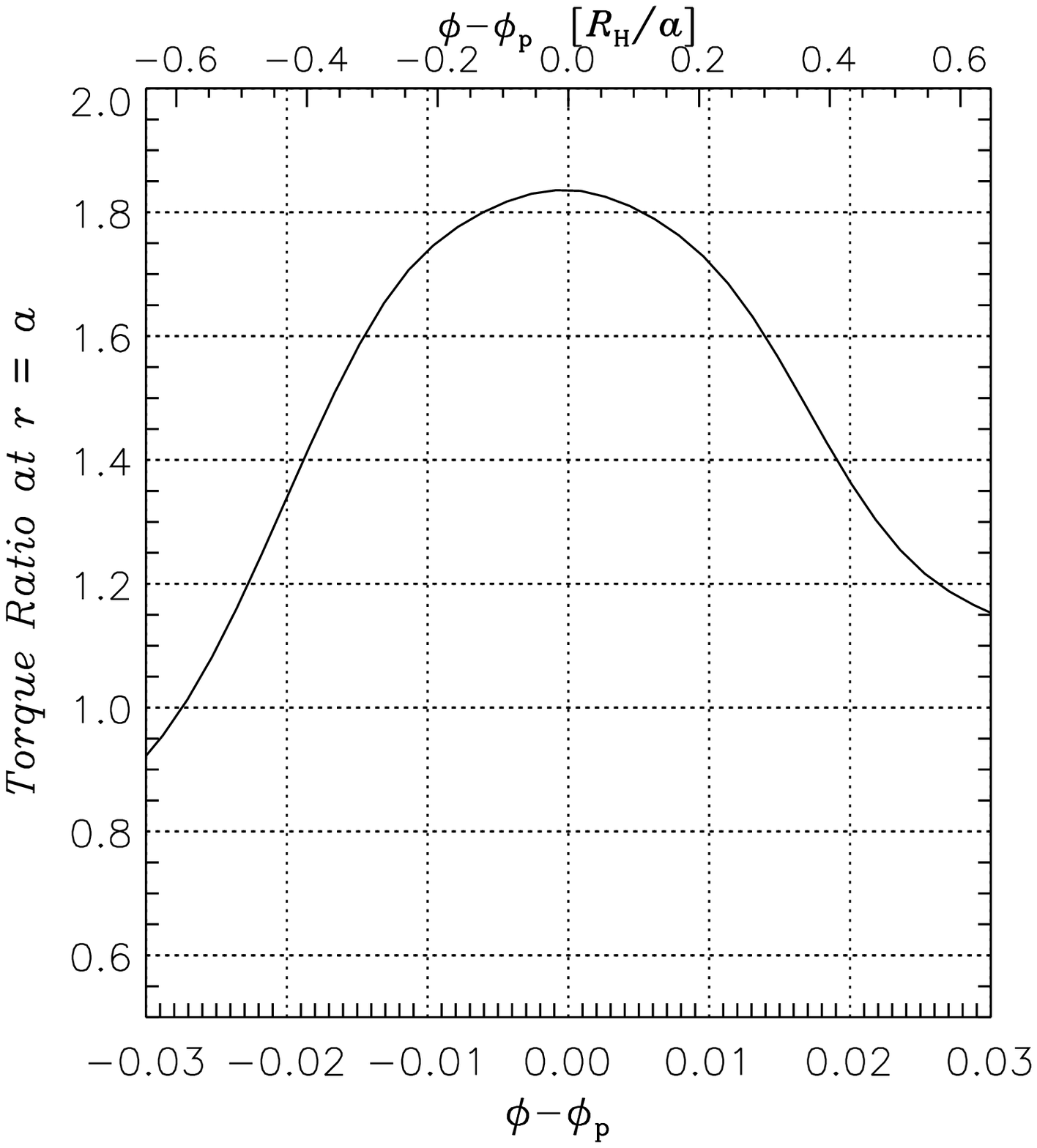}}
\caption{\small%
         \textit{Top}. Absolute value of the torque density (logarithmic scale)
         and contour lines around and
         inside the Hill sphere of a planet with $\Mp/\MStar=3\times10^{-4}$,
         orbiting in a high-mass disc.
         The snapshots are taken at $t=537$ orbital periods, i.e., after the 
         planet has migrated for $60$ orbits. The left panel refers to the 
         calculation executed with the grid system 2D4Gb while the right panel
         refers to that executed with the grid system 2D5Gb.
         The torque density is negative when $\phi<\phi_{\rmn{p}}$ and positive
         when $\phi>\phi_{\rmn{p}}$.
         Note that the torque density at the planet position is not exactly
         zero because the planet does not sit on the centre of a mesh zone,
         where density torque is computed.
         \textit{Bottom}. The left panel shows the cut of the torque density 
         magnitude through the planet's radial position for the models shown
         in the top panels. The solid line represents the outcome of the
         higher resolution simulation (2D5Gb) whereas the dashed line
         corresponds to the lower resolution simulation (2D4Gb). 
         The ratio between these two curves (higher to lower resolution
         calculation) is displayed in the right panel. The profile appears
         asymmetric (with respect to $\phi-\phi_{\rmn{p}}=0$) only towards
         the outer regions $|\phi-\phi_{\rmn{p}}|>0.02\simeq 0.4\,\Rhill/a$.
         }
\label{fig:map_torden}
\end{figure*}
The reason for the very fast migration rate, measured at the lowest 
resolution (single-level grid 2D1Gb), can be understood by examining 
the two dimensional linear map of the torque density magnitude in the 
left panel of \refFgt{fig:map_tordenst}. 
The plot describes the situation after $19$ orbits from the planet's
release, when it is migrating inwards at an average rate of roughly
$10^{-2}\,\AU\,\rmn{yr}^{-1}$.
The map clearly shows how the poor resolution (the grid zone size is
indicated by the shaded pixels) cannot properly handle
the large torque gradients within the Hill sphere and produces a very
large differential torque. This resolution effect led to the vastly 
different migration time-scales between the lowest and highest curves in
the left panel of \refFgt{fig:map_acomp}. 
A cut of the torque density magnitude, through the planet's radial 
position, is shown in the right panel of \refFgt{fig:map_tordenst} 
for both the computation executed with the grid 2D1Gb (solid line) 
and that executed with the high-resolution grid system 2D5Gb (dashed line).
The dashed-line profile was rescaled so that the maximum values were 
similar to those of the solid-line profile. The filled circles
represent the actual data.
The low-resolution torque density is highly asymmetric. The two maxima
alone exert a negative torque that would result in a migration time-scale 
of $80$ orbits.
The large mismatch between the torque density extrema is not observed in 
high-resolution model, in which their opposite sign contributions
nearly cancel each other.

The differences between the two highest resolution calculations discussed
here are less evident and require some discussion.
Two-dimensional logarithmic maps of the magnitude of the torque density 
for such models are shown in the top panels of \refFgt{fig:map_torden}. 
They were obtained from the computations with the grid systems 2D4Gb (left) 
and 2D5Gb (right).
Both maps describe the situation $60$ orbits after the planet's release.
The torque density is positive on the side leading the planet, 
$\phi>\phi_{\rmn{p}}$, and negative on the opposite side 
($\phi<\phi_{\rmn{p}}$).
As clearly indicated in the Figure,
the torque density within the inner half of the Hill sphere is orders
of magnitudes larger than it is anywhere else in the surrounding region and,
therefore, in the whole disc. 
This is the reason why the migration speed is so susceptible to the torques
exerted within the planet's Hill lobe.
Any mismatch between the positive ($\phi>\phi_{\rmn{p}}$) and negative 
($\phi>\phi_{\rmn{p}}$) contributions can produce a very large net 
(either positive or negative) torque acting on the planet. 
From the top panels in \refFgt{fig:map_torden},
the torque density magnitude appears rather symmetric with respect to the 
direction $\phi=\phi_{\rmn{p}}$. This is clear
from the bottom-left panel, where cuts through the planet's radial position 
are compared for the two grid systems. The solid line corresponds to the
more resolved model.
Nevertheless, the results shown in the bottom-left panel of 
\refFgt{fig:map_acomp} 
imply that the torque exerted by the Hill sphere is more positive 
(i.e., greater than zero and larger) in the higher resolution model 
(grid system 2D5Gb) than it is in the lower resolution one 
(grid system 2D4Gb).
Indeed, this effect is highlighted by the ratio between 
the two torque density cuts (solid to dashed profile, that is higher to lower
resolution results) reported in the bottom-right panel of 
\refFgt{fig:map_torden}.
The important thing to note is that the curve is asymmetric, with respect to 
the direction $\phi=\phi_{\rmn{p}}$, towards the the outer parts of the 
Hill sphere $|\phi-\phi_{\rmn{p}}|> 0.02\simeq 0.4\,\Rhill/a$.
This means that the mismatch between the positive ($\phi>\phi_{\rmn{p}}$) 
and negative ($\phi<\phi_{\rmn{p}}$) torques arising from the region
$|\phi-\phi_{\rmn{p}}|> 0.02$
produces a net positive torque that is greater in the higher resolution 
model than it is in the lower resolution model.  
Most of the asymmetry, and therefore the discrepancy between the two 
computations, must be confined to the region enclosed between roughly 
$0.4\,\Rhill$ and $0.75\,\Rhill$ 
from the planet because convergence tests executed with the configuration 
$\beta=0.75$ gave the same migration behaviour for the two models.

We also performed 3D simulations, using the grid system 3D3Gb (see 
\refTab{tbl:grids2}), yet no appreciable differences from 2D calculations
executed with the grid system 2D3Gb were observed. This was easily 
predictable, given the large smoothing length adopted in these models and 
the very small the disc thickness.

\subsubsection[]{A converged migration rate}
\label{sec:map_conv}
\begin{figure}
\centering%
\resizebox{1.0\linewidth}{!}{%
\includegraphics{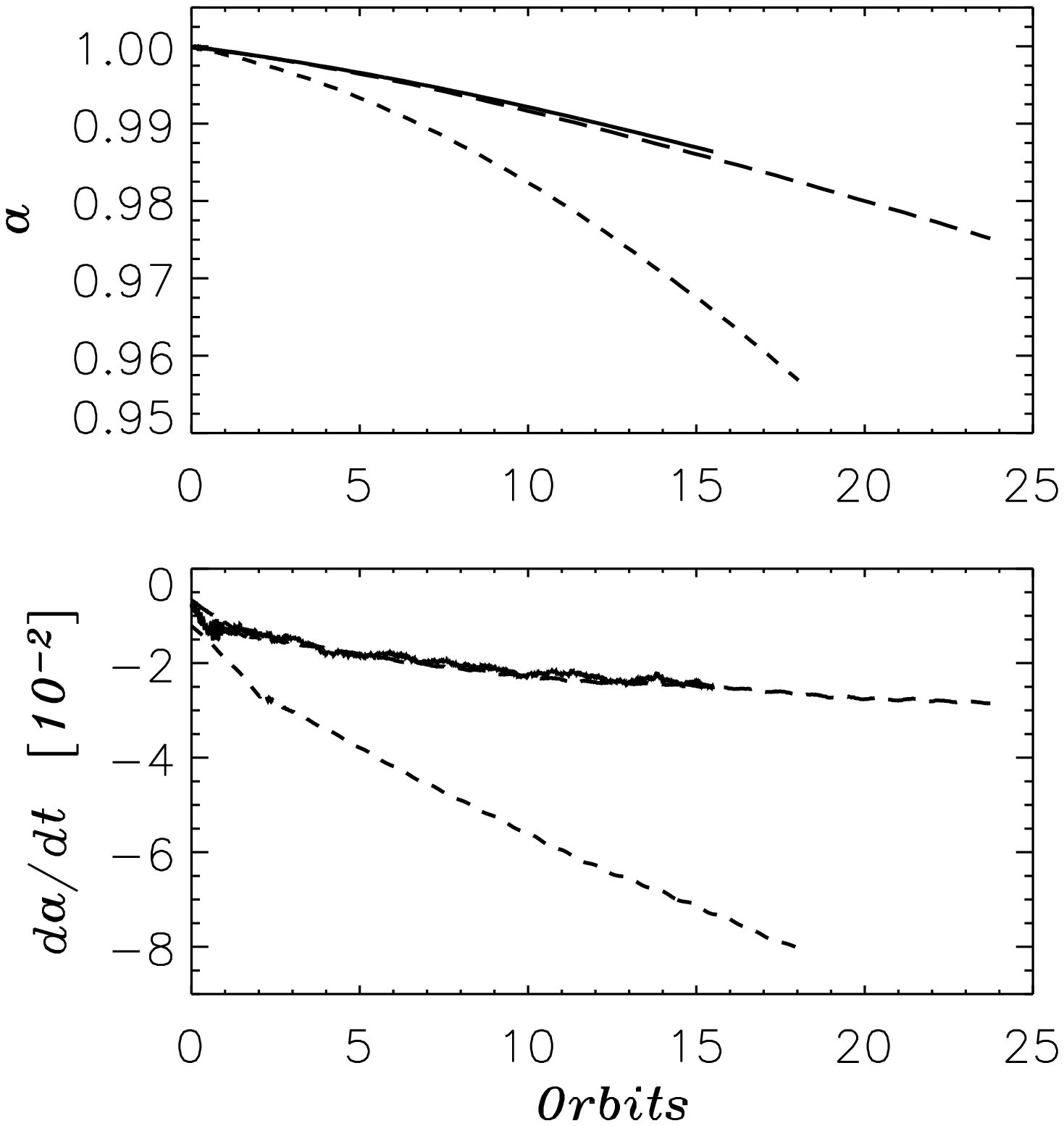}}
\caption{\small%
         Semi-major axis evolution (\textit{top}) and migration
         speed  (\textit{bottom}) obtained from the grid system
         2D6Gb (solid line) versus those calculated
         with grid systems 2D5Gb (long-dash line)
         and  2D4Gb (short-dash line). The units of $\dot{a}$
         are initial Hill radii per orbit.
         In these simulations the release time was equal to
         $t_{\rmn{rls}}=277$ orbits (see text).
         All torques are taken into account (i.e., $\beta=0$).
         This additional convergence test shows 
         that the migration behaviour given by the
         grid system 2D5Gb is fundamentally a converged
         one.
        }
\label{fig:map_conv}
\end{figure}
In order to evaluate how close to convergence the orbital evolution 
given by the grid system 2D5Gb is (see \refFgp{fig:map_acomp}, left 
panel), we made a final attempt and ran a model with the grid system
2D6Gb (see \refTab{tbl:grids2}), which resolves the Hill radius with 
about $104$ grid zones. However, we could not run a complete model as 
those in \refSect{sec:hm_ct}.
In fact, evolving a model for about $550$ orbits with such a grid 
system would have required around $8000$ CPU hours. 
We only had the computational resources to run this particular 
model for about $292$ orbital periods. Therefore we set the 
release time to $t_{\rmn{rls}}=277$ orbits and let the planet 
migrate for about $15$ orbits. 

To carry out a consistent comparison, we performed calculations
with the grid systems 2D4Gb and 2D5Gb imposing the same release 
time. 
The results are shown in \refFgt{fig:map_conv}. Despite the short 
time over which the planet actually migrated, the highest 
resolution model (solid lines) provided evolutions of both $a$ 
(top panel) and $\dot{a}$ (bottom panel) that are in very good 
agreement with those computed with the grid system 2D5Gb 
(long-dash lines).
This implies that the rates of migration obtained with the latter 
grid system (2D5Gb) can be considered as converged rates.
This also indicates that in order to accurately compute torques 
from within and around the Hill sphere, in a configuration as
described in \refSect{sec:hm_par}, linear resolutions on 
the order of $52$ grid zones per Hill radius are required.

It has to be emphasised that, while for Jupiter-mass planets in 
low-mass discs torques were converged with respect to both
the numerical resolution and the smoothing length of the planet's
potential (see \refSecp{sec:lm_models}), in the present case we 
only examined convergence with respect to the numerical resolution.

\subsection[]{Comparison of migration rates of static and migrating 
planets: the Saturn-mass case}
\label{sec:msat_hmd}
We performed the same type of comparison, as done above, for the torques 
acting on a static and migrating planet. We considered both the 
configurations $\beta=0$ and $\beta=1.0$. These results were obtained 
from the calculation run with the grid system 2D4Gb ($26$ grid zones
per Hill radius).
Recall that with $\beta=1.0$, complete numerical convergence 
was attained with $13$ grid zones per Hill radius and therefore 
the same result is produced by the model executed with the grid system 
2D5Gb. With $\beta=0$, convergence was presumably obtained only with
this last grid system. Nonetheless, it is still of interest to 
investigate if migration times-scales depend on whether the planet 
is allowed to migrate or kept on a fixed orbit, with the grid system 
2D4Gb, since it yields a larger migration rate.

Using the torques measured during the last ten orbits before the planet
is released, by the same procedure outlined in
\refSect{sec:mjup_lmd}, we obtained static migration time-scales of 
$\taum^{\rmn{S}}=770$ initial orbits for $\beta=0$ and 
$\taum^{\rmn{S}}=540$ initial orbits for $\beta=1.0$. 
Both values are remarkably close to those obtained when the planet is 
allowed to move: $\taum=773$ and $\taum=493$ initial orbits, respectively 
(see \refFgp{fig:map_acomp}).

\section[]{Discussion and Conclusions}
\label{sec:discussion}
We calculated the migration rates of planets embedded in discs.
These calculations were performed in two and three dimensions, using
a reference frame that corotates with the planet and
a nested-grid code that can provide high resolution close to the planet
while it migrates.
The models span a variety of smoothing parameters for the potential
and a variety of grid resolutions. Both accreting and non-accreting 
boundary conditions near the planet were considered.
We were especially interested in whether torques from the coorbital 
region can lead to runaway migration, as reported in MP03.

In the case of a Jupiter-mass planet embedded a low-mass disc
($\Md=0.01\,\MSun$ within $26\,\AU$), the planet opens a gap in
which there is some flowing material.
Numerical convergence was readily obtained (see \refFgp{fig:jup_acomp}).
The torques arising from within the Hill sphere do make a contribution
to the torque (up to $60$ per cent), but always in the sense of reducing
the migration rate. The migration time-scales are
numerically of order the Type~II migration time-scale
$2\,a^2/(3\,\nu) \simeq 10^4$ orbits (see \refTab{tbl:static_moving}) 
and much longer than the Type~I time-scale of about $5\times 10^{2}$ 
orbits. 

In the case of a Saturn-mass planet embedded in a high-mass disc 
($\Md=0.02\,\MSun$ within $13\,\AU$), the planet opens a less clean gap 
and is much more susceptible to the larger amount of material that 
resides in the coorbital region.
Numerical convergence in this case was much more difficult to achieve
when torques from within the Hill sphere were included.
Convergence was more easily obtained
when considering only torques from outside the Hill sphere
(see \refFgp{fig:map_acomp}).
The reason is that the mass of gas flowing within the Hill sphere is
larger than the mass of the planet. Any inaccuracies in the density 
structure near the planet (e.g., due to finite resolution) can lead 
to strong net torques (see \refFgp{fig:map_tordenst} and  
\refFgp{fig:map_torden}).
Although numerically converged, migration rates that do not account
for torques from within the Hill sphere (or a large fraction of it)
are artificially large (compare curves for grid system 2D5Gb in the 
left and right panels of \refFgp{fig:map_acomp}). This also affects
the flow structure around the planet and in the coorbital region
(see \refFgp{fig:map_denstream}).
With increasing resolution, the gas mass within the Hill sphere increases, 
yet the migration rate decreases. 
In the case that the resolution was about equal to the smoothing length, 
which was $0.39\,\Rhill$, the migration rate was very high, comparable 
to the Type~I migration rate (as also found by MP03). However at the 
highest resolution we applied with a release time of $477$ orbits, 
which was sixteen times higher (in terms 
of linear resolution), the migration rate dropped dramatically by more 
than two orders-of-magnitude. 
Discrepancies between higher resolution calculations are more subtle and
arise from the outer half of the Hill sphere (see \refFgp{fig:map_torden}).
A calculation based on the grid system 2D6Gb, for which 
$\Rhill/\Delta r=104$, indicates that accurately describing torques from 
around and inside the Hill sphere requires resolutions of at least
$52$ grid zones per Hill radius.
At highest resolution, the migration time-scale is about $3\times10^3$ 
orbits (see left panel of \refFgp{fig:map_acomp}), 
somewhat shorter than the Type~II time-scale. 
But the process is unlikely to be simply described by Type~II migration.

We calculated the torques exerted by the disc on planets whose orbits 
are fixed and used these to obtain migration time-scales. Comparing these 
time-scales to those obtained by releasing the planet and allowing it to
migrate freely through the disc, we found no significant difference in the 
migration time-scales. This argues that the corotation torques are not
greatly affected by the radial drift of a planet.

In summary, the migration rates for planets that open impure gaps (in 
which some material flows) are substantially smaller than Type~I rates 
and do not seem to be simply described by Type~II migration.
Torques arising near the
planet can be important, but do not appear to have a dramatic effect
in raising the rates.
Resolution is key to obtaining accurate torques.
\section*{Acknowledgments}
We thank the referee, P.\ Armitage, for his prompt and useful comments.
We also thank F.\ Masset and J.\ Papaloizou for carefully reading the
manuscript and providing comments.
The computations reported in this paper were performed using 
the UK Astrophysical Fluids Facility (UKAFF).
GD is grateful to the Leverhulme Trust for support under a UKAFF 
Fellowship and acknowledges support from the STScI Visitors Program.
SL acknowledges support from NASA Origins of Solar Systems grants 
NAG5-10732 and NNG04GG50G.


\appendix
\begin{twocolumn}
\section[]{Numerical tests}
\label{ap_sec:num_tests}
The purpose of this Appendix is to demonstrate the reliability of the 
nested-grid technique when it is applied to disc-planet interaction 
calculations and, more specifically, to planetary migration.
The capabilities of this technique in the context of astrophysical 
fluid-dynamics modelling have been addressed by a number of authors 
\citep*[e.g.,][and references therein]%
{ruffert1992,yorke1993,yorke1995,ziegler1997}.
We therefore concentrate on specific test computations that closely 
concern the application we did in this paper.
The tests that we report here were done in two dimensions only to avoid 
excessively long computing times.

We set up a model of a Jupiter-mass planet ($\Mp/\MStar=10^{-3}$) in 
a massive disc with $\Md=1.1\times10^{-2}\,\MStar$ (inside $13\,\AU$
of a $1\,\MSun$ star) and whose aspect ratio is $H/r=0.03$.
The initial surface density drops as $r^{-1/2}$ but it also includes a 
theoretical gap along the planetary orbit, as done for most of the 
Jupiter-mass models discussed so far.
The adopted viscosity prescription was the same as that chosen for all
the other calculations (see \refSecp{sec:lm_par}). 
The radial extent of the computational domain ranges from $0.4$ to $2.5$
length units and reflective boundary conditions were applied to both 
edges of the disc to enforce mass conservation (the planet was not 
accreting). The planet's orbit was initialised with $a_{0}=1$ (and with 
zero eccentricity) and it was kept steady for the first orbital period 
(i.e., $t_{\rmn{rls}}=1$). 
The reference frame was set to rotate at a variable rate
$\Omega=\Omega(t)$ so that $\phi_{\rmn{p}}=\pi$ throughout the simulations, 
according to the procedure introduced in \refSect{sec:rot_elements}.

\begin{figure}
\centering%
\resizebox{1.0\linewidth}{!}{%
\includegraphics{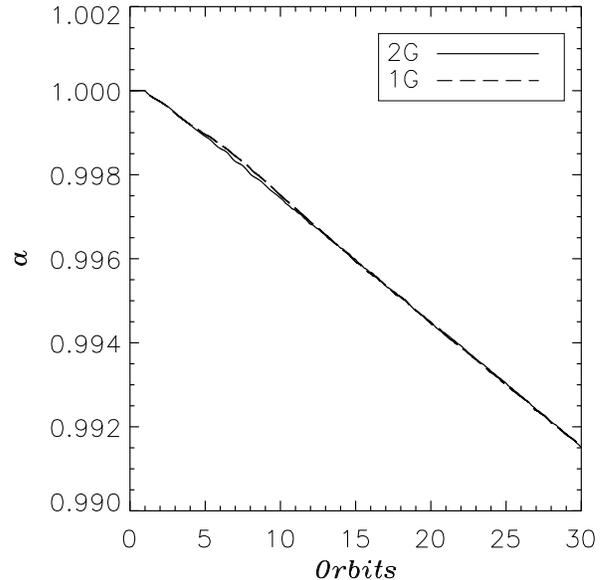}}
\caption{\small%
         Semi-major axis evolution of the test model as calculated
         in a nested-grid mode with the grid system 2G (solid line) 
         and in a single-grid mode with the grid 1G (dashed line). 
         Radial and azimuthal resolutions 
         ($\Delta r\simeq\Delta\phi\simeq 7.0\times10^{-3}$) 
         coincide over roughly a half of the whole computational 
         domain ($[0.4,2.5]\times 2\,\pi$).
         }
\label{fig:test-2g_vs_1g-uno}
\end{figure}
\begin{figure}
\centering%
\resizebox{1.0\linewidth}{!}{%
\includegraphics{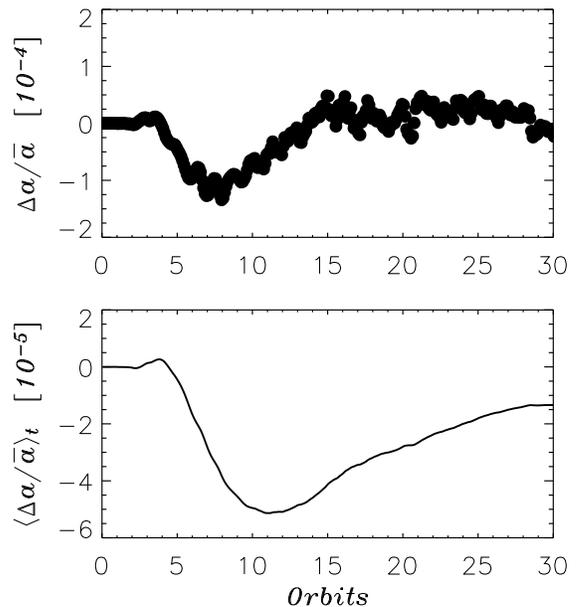}}
\caption{\small%
         \textit{Top}. 
         Normalised difference of the two functions $a_{\rmn{2G}}$
         and $a_{\rmn{1G}}$, shown in \refFgt{fig:test-2g_vs_1g-uno},
         according to \refeqt{eq:adiff}.
         Each filled circle represents the average of the data over 
         time-intervals of a tenth of an orbit that are centred at 
         the middle point of each time-interval.
      \textit{Bottom}. 
         Running time average of the data displayed in the top panel,
         according to \refeqt{eq:arunav}.
         }
\label{fig:test-2g_vs_1g-due}
\end{figure}
In order to evaluate in quantitative terms the behaviour of the 
nested-grid technique, the model outlined above was executed with a 
two-level grid system (i.e., in nested-grid mode) as well as with a 
single-level grid (i.e., in single-grid mode).
The first grid level of the two-level grid system (henceforth 2G), 
which covers the whole computational domain, had 
$N_{r}\times N_{\phi}=147\times455$ grid zones
($\Delta r\simeq 1.46\times10^{-2}$ and $\Delta\phi\simeq 1.40\times10^{-2}$), 
whereas the second level had $N_{r}\times N_{\phi}=264\times464$ zones
($\Delta r\simeq 7.3\times10^{-3}$ and $\Delta\phi\simeq 7.0\times10^{-3}$).
With this setup, the higher resolution region extends from $r\simeq0.5$
to $r\simeq2.4$ and from $\phi\simeq\pi/2$ to $\phi\simeq 3\,\pi/2$.
In order to achieve the same resolution with a single-level grid
(henceforth 1G), the mesh must have $N_{r}\times N_{\phi}=290\times904$ 
grid zones.
Although the grid system 2G covers nearly a half of the whole domain
with the same numerical resolution as the grid 1G, the perturbations
induced by the planet propagate to the entire disc over a short
time-scale and after $5$ orbits the spiral wave pattern has already 
developed.
Therefore, after a few orbits, one should expect that the results of
the two simulations start to differ. The discrepancy depends upon the
ability of the first level of the grid 2G to capture the same flow
features as the grid 1G does, in the region $\phi<\pi/2$ and 
$\phi>3\,\pi/2$.

Since this study is about migration, we focused on the evaluation and
comparison of the semi-major axis evolutions, which also give a direct 
indication of the acting torques as a function of time.
Note that this is a more strict test than simply comparing the torque 
distributions at certain times, which would \textit{only} imply that 
$\dot{a}$ is the same at those times.
The orbital decay for the two simulations is shown in 
\refFgt{fig:test-2g_vs_1g-uno}. The solid and dashed lines pertain to 
grids 2G and 1G, respectively.
To estimate the differences in more detail, we computed the normalised 
difference
\begin{equation}
   \frac{\Delta a}{\bar{a}} =2\left(\frac{a_{\rmn{II}}-a_{\rmn{I}}}%
                     {a_{\rmn{II}}+a_{\rmn{I}}}\right),
\label{eq:adiff}
\end{equation}
where the labels $\rmn{I}$ and $\rmn{II}$ identify the used grids 1G 
and 2G, respectively.
Since the time-step is different in the two simulations, $a_{\rmn{1G}}$
and $a_{\rmn{2G}}$ were averaged over time-intervals of $0.1$ orbits and 
the value measured from \refeqt{eq:adiff} was assigned to the central time
of each interval. As shown in the top panel of \refFgt{fig:test-2g_vs_1g-due},
$\Delta a/\bar{a}$ is typically a few times $10^{-5}$ and it doesn't 
increase beyond $1.5\times10^{-4}$.
The bottom panel of \refFgt{fig:test-2g_vs_1g-due} illustrates the running
time average of the normalised difference
\begin{equation}
   \left\langle\frac{\Delta a}{\bar{a}}\right\rangle_{t}=%
   \frac{1}{t}\int_0^{t}\frac{\Delta a}{\bar{a}}\,\rmn{d}t',
\label{eq:arunav}
\end{equation}
which indicates that the discrepancies in the orbital evolutions are on 
average around a few times $10^{-3}$ per cent.

From the viewpoint of the computational load, the advantage of the 
nested-grid technique is remarkable: in the computations reported above, 
the time-step required for numerical stability by grid 2G (first level) 
is twice as long as that required by grid 1G and so the time needed to 
complete one orbital period is twice as short. 
It is also worthwhile to point out that the refinement capabilities of 
the nested-grid strategy does not reduce the accuracy of the numerical 
algorithm, which strictly remains second-order accurate in space since 
the mesh step size is always constant on each grid level.

\begin{figure}[t!]
\centering%
\resizebox{1.0\linewidth}{!}{%
\includegraphics{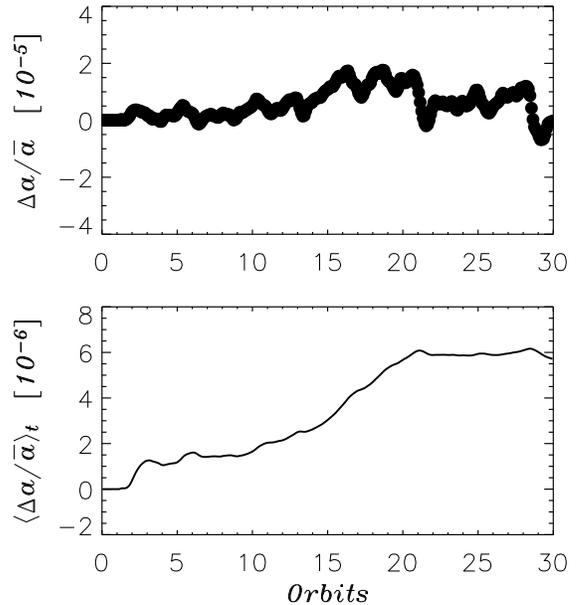}}
\caption{\small%
         Comparison between the orbital evolutions computed with the 
         grid 1G setting a constant rotation rate 
         $\Omega=\Omega_{0}=\sqrt{G\,\MStar/a^{3}_{0}}$ 
         and a variable rate (i.e., $\dot{\Omega}\neq0$) such that
         the planet's azimuthal position remains constant.
         \textit{Top}. 
         Normalised difference (see \refeqp{eq:adiff}) sampled
         as in the top panel of \refFgt{fig:test-2g_vs_1g-due}.
         \textit{Bottom}. 
         Running time average of the data displayed in the top panel.
         }
\label{fig:test-acc_vs_rot}
\end{figure}
For the sake of completeness, we show a test on the accelerated grid technique 
that we implemented and employed in this work. Other tests (not reported here) 
on the angular momentum conservation of both the disc and the planet proved 
that conservation was achieved down to the machine precision.
Thus, we refer to a more relevant situation and show how the migration 
calculated in a reference frame rotating with a variable $\Omega$ compares 
to that calculated in a uniformly rotating grid with 
$\Omega=\Omega_{0}=\sqrt{G\,\MStar/a^{3}_{0}}$.  
The outcome of such comparison is illustrated in \refFgt{fig:test-acc_vs_rot}.
We computed the normalised difference (top panel) and the running time average 
(bottom panel), where the quantity $a_{\rmn{II}}$ in \refeqt{eq:adiff} was 
obtained from the grid 1G with $\dot{\Omega}\neq0$ while the quantity 
$a_{\rmn{I}}$ was obtained from the grid 1G with static rotation (i.e., 
$\dot{\Omega}=0$). As \refFgt{fig:test-acc_vs_rot} proves, the discrepancy 
between the two semi-major axis evolutions is not significant.
It has to be stressed that although the planet moves on a ``continuous''
path, the gravitational potential is centred in a grid zone. 
Therefore, results \textit{cannot} be exactly the same since the planet's 
trajectory through the grid centres is different in the two simulations.
In fact, in the model with  $\dot{\Omega}\neq0$ there is only radial
motion due to migration, thus the time taken by the planet to cross a
grid zone is $\Delta r/\dot{a}\simeq \left(\Delta r/a\right)\,\taum$.
Instead, in the other model, the azimuthal drift soon becomes the 
fastest component of the planet's motion and the time needed to cross
a grid zone is then given by 
$\Delta \phi/\left|\Omega_{\rmn{K}}-\Omega_{0}\right|$ or
$\left(\Delta \phi/2\pi\right)%
\left[1-\sqrt{\left(a/a_{0}\right)^{3}}\right]^{-1}$ orbits.
For instance, when $\Delta r/a\approx \Delta \phi$ and $a=0.99\,a_{0}$,
in a steadily rotating grid a planet undergoes 
$\approx\taum/11\approx 300$ (see \refFgp{fig:test-2g_vs_1g-uno})
as many encounters with the grid centres as it does in a grid rotating
at a variable rate where there is no azimuthal drift.
\end{twocolumn}



\end{document}